\documentclass{ws-ijmpb}

\usepackage{graphicx}
\usepackage{graphics}
\usepackage{latexsym,amsmath,amssymb,bm,euscript}
\usepackage{color}
\usepackage{dcolumn}
\usepackage{bm}
\usepackage{epstopdf}
\usepackage{times}
\usepackage{multirow}
\usepackage{subfigure}
\usepackage{bbm}
\usepackage[normalem]{ulem}

\def\ra{\rangle}
\def\la{\langle}

\def\Hc{{\rm H.c.}}

\begin{document}

\markboth{Hsin-Hua Lai}{Short-ranged interaction effects on the $Z_2$ topological phase transition}

%
\catchline{}{}{}{}{}
%

\title{Short-ranged interaction effects on $Z_2$ topological phase transitions: The perturbative mean-field method}

\author{Hsin-Hua Lai}

\address{
National High Magnetic Field Laboratory, Florida State University \\
Tallahassee, Florida 32310,\\ USA\\
lai@magnet.fsu.edu}

\author{Hsiang-Hsuan Hung}

\address{Department of Physics, The University of Texas at Austin,\\
Austin, TX 78712, USA\\
hhhung@physics.utexas.edu}

\maketitle

\begin{history}
\received{(Day Month Year)}
\revised{(Day Month Year)}
\end{history}

\begin{abstract}
Time-reversal symmetric topological insulator is a novel state of matter that a bulk-insulating state carries dissipationless spin transport along the surfaces, embedded by the $Z_2$ topological invariant. In the noninteracting limit, this exotic state has been intensively studied and explored with realistic systems, such as HgTe/(Hg,Cd)Te quantum wells. On the other hand, electronic correlation plays a significant role in many solid-state systems, which further influences topological properties and triggers topological phase transitions. Yet an interacting topological insulator is still an elusive subject, and most related analyses rely on the mean-field approximation and numerical simulations. Among the approaches, the  mean-field approximation fails to predict the topological phase transition, in particular at intermediate interaction strength without spontaneously breaking symmetry. In this review, we develop an analytical approach based on a combined perturbative and self-consistent mean-field treatment of interactions that is capable of capturing topological phase transitions beyond either method when used independently.  As an illustration of the method, we study the effects of short-ranged interactions on the $Z_2$ topological insulator phase, also known as the quantum spin Hall phase, in three generalized versions of the Kane-Mele model at half-filling on the honeycomb lattice. The results are in excellent agreement with quantum Monte Carlo calculations on the same model, and cannot be reproduced by either a perturbative treatment or a self-consistent mean-field treatment of the interactions. Our analytical approach helps to clarify how the symmetries of the one-body terms of the Hamiltonian determine whether interactions tend to stabilize or destabilize a topological phase.  Moreover, our method should be applicable to a wide class of models where topological transitions due to interactions are in principle possible, but are not correctly predicted by either perturbative or self-consistent treatments.
 \end{abstract}



\section{Introduction}
The birth of the topological insulators (TI) in recent years has been one of the most exciting events in condensed matter and material science fields due to their novelty and potential technological applications\cite{fu2007,moore2007,moore2010,roy2009,hasan2010,qi2011}.  Shortly after the theoretical prediction\cite{Bernevig2006}, the first experimental realization of a time-reversal symmetry (TRS) protected TI was reported in HgTe/(Hg,Cd)Te quantum wells\cite{Konig2007,Roth:sci09}. Though in all the accepted experimental examples of TI to date, the presence of the topological state and most of its properties can be well understood within a noninteracting model, it is, however, generally believed that interactions can lead to qualitatively new topological phenomena in both two\cite{Young:prb08,Neupert:prb11,Qi11,Levin:prl09,Maciejko:prb13,Ruegg:prl12,Miguel2013} and three dimensions\cite{Kargarian:prl13,Maciejko:prl14,Pesin:np10,Kargarian:prb11,Maciejko:prl10,Wan:prb11,Go:prl12}.  In two-dimensions, the Kane-Mele (KM) model\cite{kane2005a} has played an especially important role in the study of  $Z_2$ TI [also known as quantum spin Hall  insulators  (QSH)]. The KM model consists of two time-reversed copies of the Haldane model\cite{haldane1988} on a two-dimensional (2D) honeycomb lattice, with real first-neighbor hopping and imaginary second-neighbor hopping arising from spin-orbit coupling (SOC). To study interaction effects, the KM model has been supplemented with an onsite Hubbard $U$-term
\begin{equation}
\label{eq:H_U}
H_U=U \sum_i n_{i\uparrow} n_{i\downarrow},
\end{equation}
where $n_{i \sigma}$ is the number of electrons on site $i$ with spin $\sigma$. The so called Kane-Mele-Hubbard (KMH) model is  investigated extensively\cite{rachel2010,yu2011,zheng2011,hohenadler2011,Budich:prb12,wuwei2012,hohenadler2012,Griset2012,lang2013}, particularly with quantum Monte Carlo (QMC) which is free of the fermion sign problem\cite{zheng2011,hohenadler2011,hohenadler2012,lang2013}.  Its phase diagram is now well understood. Beyond the critical value of interaction strength $U_c$, there exists a magnetic phase transition which turns a topological state to an easy-plane antiferromagnetic order state.

Recently, several fermion sign-free extensions of the KMH model have been proposed and studied with QMC\cite{Hung2013,hung2014} with the goal of understanding short-ranged interaction effects on the hopping-parameter-driven $Z_2$ topological phase transitions at half-filling and at intermediate interaction, i.e. $U<U_c$. The numerical results concluded that the onsite Hubbard interaction leads either to stabilize the QSH against the parameter that drives the $Z_2$ topological phase transition from a topological insulating phase to a trivial insulating phase or to destabilize it by making it more fragile to the parameter. Though the $Z_2$ topological phase transitions have been examined using QMC, there has been no proposal of  an ``analytically`` physical picture that captures the numerical results. Most important of all, without symmetry breaking, the conventional mean-field approximation in some cases does not provide any correction to realize the topological boundary shifts observed in QMC. In this brief review article, we propose an analytical method, dubbed perturbative mean-field (PMF) method, which combines the perturbative treatment and the mean-field treatment. For pedagogy and illustrating the power of the general analytical framework we introduce, we study three variants of KMH--generalized Kane-Mele model (GKM), dimerized Kane-Mele model (DKM), and stagger-potential Kane-Mele model (SKM) described by Eq.~\eqref{eq:H_G}, Eq.~\eqref{eq:H_D} and Eq.~\eqref{eq:H_S}, supplemented by a Hubbard-$U$ term Eq. (\ref{eq:H_U}).

To study the effects of a ``local`` interaction, the conventional mean-field treatment is to directly mean-field decouple the local interaction as
\begin{eqnarray}
U\sum_i n_{i \uparrow} n_{i \downarrow}   \simeq  \frac{U}{2} \sum_{{\bf r},a=A,B} \bigg{[} \left\la n( {\bf r},a) \right\ra n({\bf r},a) + \left\la s_z ({\bf r},a) \right\ra s_z ({\bf r},a) \bigg{]},~~ \label{Eq: Hubbard_1st_order}
 \end{eqnarray}
 where $n\equiv n_\uparrow + n_\downarrow$, and $s_z \equiv n_\uparrow - n_\downarrow$. $\langle n\rangle$ and $\langle s_z\rangle$ are the order parameters to describe the charge-density-wave and spin-density wave orders, where we have explicitly neglected the constant $\la n_j \ra^2$ appearing in the mean-field Hamiltonian since it only shifts the total energy. It is clear that the mean-field decoupled Hamiltonian only gives an overall density correction which renormalizes the chemical potential without renormalizing the bare hopping amplitudes. For the SKM described by Eq. ~\eqref{eq:H_S}, we will see that the usual conventional mean-field treatment is enough to predict the shift of the critical value of the relevant parameter. However, for GKM and DKM described by Eq.~\eqref{eq:H_G} and Eq.~\eqref{eq:H_D}, the relevant parameters that drive the topological transitions are the hopping amplitudes and the QMC explicitly sees that the critical values of the relevant hopping amplitudes shift in the presence of a ``local`` interaction, which can not be captured within the conventional mean-field picture. Hence, an analytical method that is beyond the conventional mean-field treatment is needed.
 
In order to develop an analytical method which can generate terms that renormalize the hopping amplitudes from a ``local`` interaction $U$, and, somehow, reproduce the results of the conventional mean-field treatment, we consider to expand the interacting term in series of $U/t$ for the partition function of the full interacting Hamiltonian, $H=H_0 + H_U$, which mimics the way of doing perturbative analysis. The expansion to first order, $O(U/t)$, gives the term of the local interaction and under mean-field decoupling we can regain the results of the conventional mean-field treatment after re-exponentiating the term to be combined with the noninteracting Hamiltonian $H_0$. The expansion to second order of $U/t $, i.e. $O(U/t)^2$, gives eight-fermion terms consisting of fermions at two different sublattices $A$ and $B$ of the honeycomb lattice. The mean-field decoupling of the second-order terms give bilinear terms consisting of the total density at site $j$ $\langle n_j \rangle$ and two-point correlation functions between sites $i$ and $j$, $\chi^*_{ij} \equiv \la c_i^\dagger c_j \ra$. Again we can re-exponentiate the terms to be combined with the original noninteracting Hamiltonian $H_0$ and examine how the bare Hamiltonian gets renormalized. During the mean-field decouplings of the first-order and second-order terms, we only keep the connected and irreducible terms which can be related to the proper self-energies in the weak-coupling perturbative calculations.\cite{Fetter-Walecka} We call this method the perturbative mean-field method (PMF). The schematic mathematical expression is 

\begin{eqnarray}
\nonumber \mathcal{Z} &=& \int \mathcal{D}[\psi^\dagger,\psi] e^{-H} = \int \mathcal{D}[\psi^\dagger,\psi] e^{-H_0}\left( 1 - H_U + \frac{1}{2}H_U^2 -\cdots \right)\\
 \nonumber &\simeq& \int \mathcal{D} [ \psi^\dagger, \psi] e^{-H_0} \left( 1 - \la H_U \ra_{MF} + \frac{1}{2} \la H_U^2 \ra_{MF} - \cdots \right)\\
&\simeq& \int \mathcal{D} [ \psi^\dagger,\psi] e^{-\left[ H_0 + \la H_U\ra_{MF} - \frac{1}{2}\la H_U^2\ra_{MF} + \cdots\right]},
\end{eqnarray}
where we can see the first order correction $\delta \mathcal{H}_1$ to the bare Hamiltonian is $\la H_U \ra _{MF}$ and the second order correction is $\delta \mathcal{H}_2 = -\frac{1}{2} \la H_U \ra^2_{MF}$. The correction terms $\delta \mathcal{H}_{1/2}$ are functions of variables $\chi_{ij}$ which need to be determined self-consistently. We will use the PMF method to examine the three different variants of KM models in Sec.~\ref{Sec:PMF}.

The GKM and DKM models preserve the discrete particle-hole symmetry (PHS) and therefore the stability of the topological phase can be addressed by exact QMC. For interaction strengths below the regime of magnetic instabilities, the QMC results show that the interactions produce a shift in the location of the phase boundary (opposite directions for the two models)\cite{hung2014}. On the other hand, the SKM model breaks PHS and therefore in the positive $U$ side, QMC suffers from the sign problem. Nevertheless, our PMF method is general and can be applied to the three different models above regardless of the sign of the interaction. We find that the sign of the shifts and their scaling-law with respect to the interaction strength in the three different models are accurately reproduced by the combination of perturbation theory and a self-consistent mean-field calculation, though they are \underline{not} captured by either one independently. Our results emphasize that short-range interactions can have subtle effects on the stability of topological phases, and may need to be treated by a method analogous to that we use here when other approaches are not available or desirable.

The review article is organized as follows. In Sec.~\ref{Sec:KMvariants_noninteracting} we introduce three variants of the KM models, their low-energy descriptions, and $Z_2$ topological phase transitions in the noninteracting limit. In Sec.~\ref{Sec:PMF}, we apply the method to different models followed by self-consistent numerical calculations. In Sec.~\ref{Sec:QMC} we present the exactly QMC results and show the consistency between these two approaches. In Sec.~\ref{Sec:Discussion} we include several relevant discussions, including RG analysis and a more extended $U-V$ interaction effects on the SKM models. In Sec.~\ref{Sec:Conclusion} we conclude with some future perspective of our general analytic method.

\section{Variants of the Kane-Mele model and the topological phase transitions in the noninteracting limit}\label{Sec:KMvariants_noninteracting}

\begin{figure}[t]
   \centering
   \subfigure{\includegraphics[width=1.5 in]{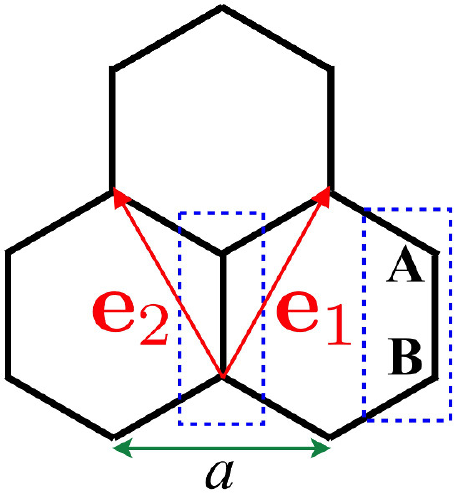}
   \label{Fig:honeycomb}}
   \subfigure{\includegraphics[width=2 in]{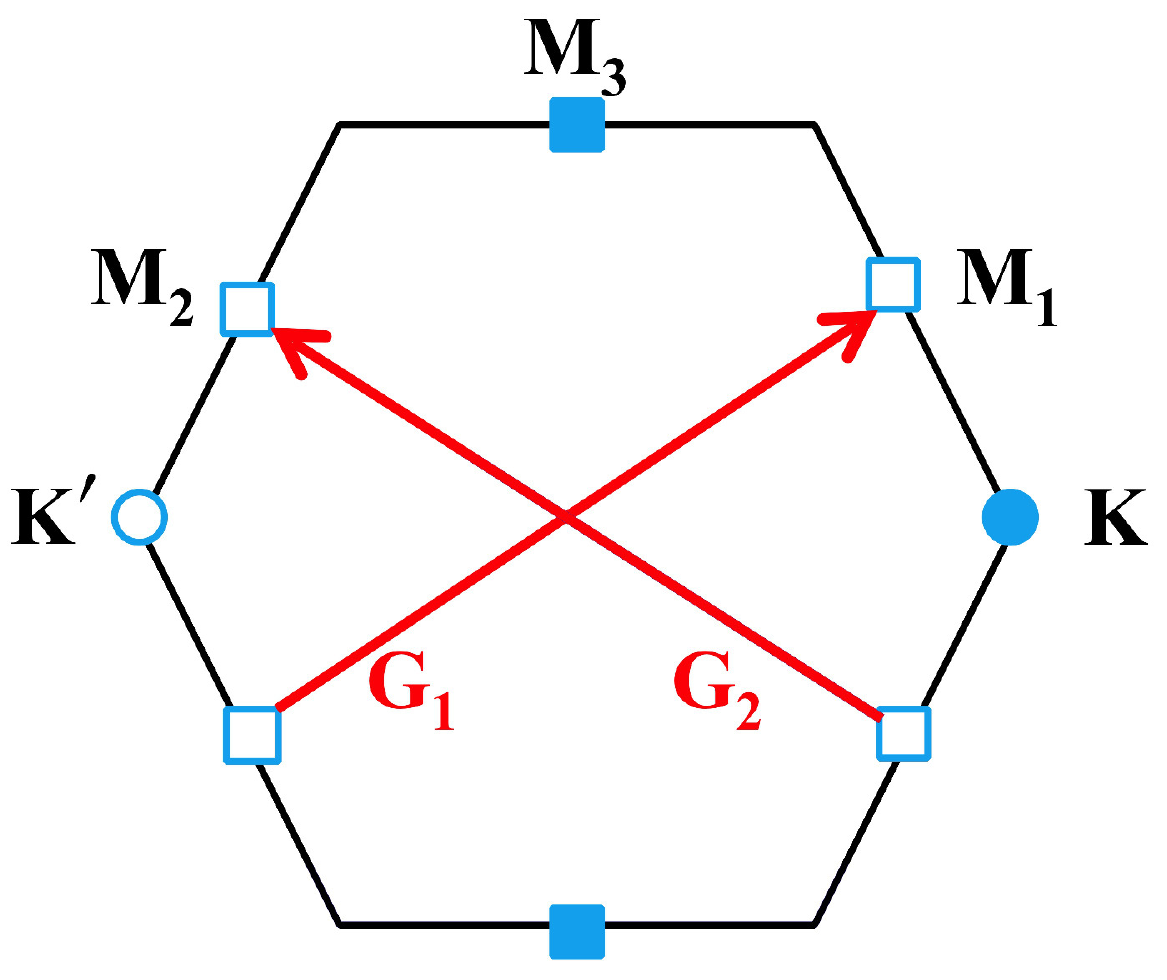}
   \label{Fig:BZ}}
 \caption{(a) Schematic of the honeycomb lattice with two sublattices labeled $A$ and $B$. The vectors ${\bf e}_{1/2} = (\pm 1/2, \sqrt{3}/2)$ connect the same sublattice in different unit cells.  The lattice constant $a$ is set to 1. (b) Illustration of the Brillouin zone (B.Z.). There are several relevant momenta that are important in the low-energy descriptions of variants of KM models--The usual momenta ${\bf K}=- {\bf K'}= (4\pi/3,0)$ as the locations of the Dirac nodes in the original KM model and the time-reversal-invariant momenta (TRIM) ${\bf M}_{1,2} \equiv (\pm \pi, \pi/\sqrt{3})$ and ${\bf M}_3\equiv (0, 2\pi/\sqrt{3})$. Note that ${\bf M}_1$ and ${\bf M}_2$ are related by $C_3$ lattice rotation symmetry. ${\bf G}_1$ and ${\bf G}_2$ are the reciprocal lattice vectors.}
\label{Fig:honeycomb+BZ}
\end{figure}

The KM model is one of early proposed models to harbor QSH. In the noninteracting limit, the original KM model consists of real-valued first-neighbor hoppings, $t$, and imaginary second neighbor hoppings, $\lambda_{so}$. The schematic of the honeycomb lattice with two sublattices ($A$ and $B$) and the corresponding relevant Brillouin zone (B.Z.) are shown in Fig.~\ref{Fig:honeycomb+BZ}.  Each spin component contributes chirality on the honeycomb lattice, to produce a nontrivial quantum anomalous Hall effect as known in the Haldane model.\cite{haldane1988} However, the opposite spin flavor carries opposite chirality, and contributes opposite sign spin Chern number $C_{\uparrow}=-C_{\downarrow}= 1$. Therefore the system is  time-reversal symmetric and total Chern number $\sum_{\sigma}C_{\sigma}=0$.
The intrinsic topological invariant of topological insulators and QSH is the $Z_2$ invariant, $\nu=1$ or $\nu=0$. With an inversion symmetry, it can be easily evaluated as\cite{fu2007}
\begin{eqnarray}
(-1)^{\nu}=\prod_{{\bf k}\in \textrm{TRIM}}\prod_n \xi_{2n}({\bf k}),
\end{eqnarray}
where $\xi_{2n}({\bf k})=\pm 1$ is the parity of the occupied eigenstates of the noninteracting Hamiltonian at TRIM points. In the KM model, the TRIM points are ${\bf \Gamma}=(0,0)$ and ${\bf M}_{1,2,3}$ as shown in Fig. \ref{Fig:BZ}. Note that due to the presence of time-reversal symmetry, $\xi_{2n-1}$ and $\xi_{2n}$ have the same parity. $\nu=1$ ($\nu=0$) denotes a nontrivial (trivial) state.

In this review article, we will focus on the short-ranged interaction effects on three variants of the KM model at half-filling on the honeycomb lattice. The three variants are what we mentioned in the Introduction section--GKM\cite{Hung2013}, DKM\cite{lang2013}, and SKM\cite{kane2005a,Lai_staggerKM}. The GKM includes real-valued third neighbor hoppings in addition to the original KM model, as illustrated in Fig.~\ref{Fig:GKM_honeycomb} while the DKM consists of anisotropic hoppings with hopping strength $t_d$ within a unit cell larger than those between different unit cells, Fig.~\ref{Fig:DKM_honeycomb}. The SKM includes staggered potentials in addition to the Hamiltonian in KM model, Fig.~\ref{Fig:SKM_honeycomb}. From the symmetry perspective, both GKM and DKM conserve the PHS, but DKM explicitly breaks the lattice $C_3$ rotation symmetry by the center of a honeycomb while GKM still preserve $C_3$. The SKM, unlike GKM and DKM, explicitly \textit{breaks} PHS even at half-filling condition due to the presence of the staggered potentials. Below we will first give the noninteracting description of each model separately followed by the discussions of the local interaction effects on the $Z_2$ topological transition.

\subsection{Generalized Kane-Mele Model}
\begin{figure}[t]
   \centering
   \subfigure{\includegraphics[width=1.5 in]{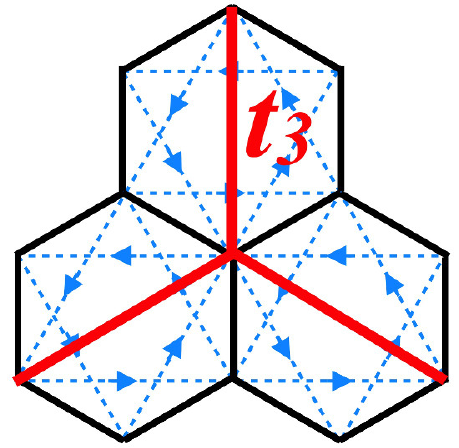}
   \label{Fig:GKM_honeycomb}}
   \subfigure{\includegraphics[width=1.8 in]{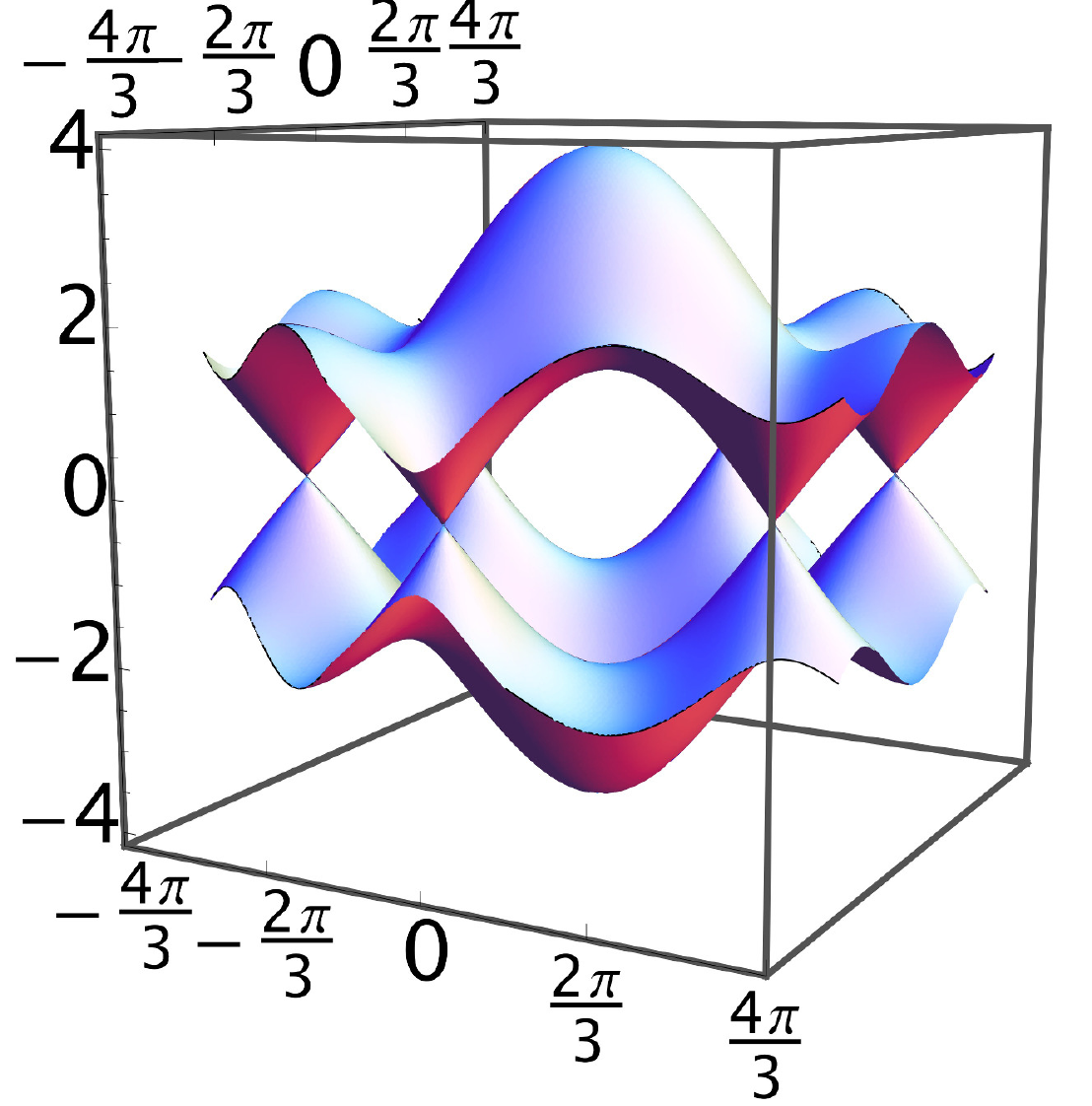}
   \label{Fig:GKM_spectrum}}
   \caption{(a) Schematic of the GKM model and (b) the band spectrums at the topological phase transition (gap closing). The blue dashed lines represent the imaginary second neighbor hoppings and the arrow directions represent their signs. The red lines represent the $t_3$ hoppings. At the phase transition point, the bands gaps close at three TRIM ${\bf M}_{a=1,2,3}$, unlike the usual KM model.}
  \label{Fig:GKM_model}
\end{figure}

The GKM Hamiltonian is
\begin{equation}
\label{eq:H_G}
H_{G}=  -\sum_{jk } \sum_{\sigma}  t_{jk} c^\dagger_{j \sigma} c_{k \sigma} + i \lambda_{so} \sum_{\la\la jk \ra\ra } \sum_{\sigma} \sigma c^\dagger_{j \sigma} \nu_{jk} c_{k\sigma},
\end{equation}
with $t_{jk} = t$ for $j, k \in \la j k \ra$, $t_{jk} = t_3$ for $j, k \in \la\la \la j k \ra \ra \ra$, and zero else, where $\la...\ra$, $\la\la...\ra\ra$, and $\la\la\la...\ra\ra\ra$ represent the nearest neighbors, the second neighbors, and the third neighbors. $\nu_{jk} = +1(-1)$ for (counter-)clockwise second-neighbor hopping and without lack of generality, we choose $t,\lambda_{so},  t_3 >0$. $\sigma=\uparrow (\downarrow) = +(-)$. The operator $c^\dagger_{j \sigma}$ ($c_{j\sigma}$) creates (annihilates) an electron at site $j$ with spin $\sigma$. The schematic is shown in Fig.~\ref{Fig:GKM_honeycomb}. 
For clarity, from here forward we replace the site labeling $j$ with $j = \{{\bf r}, a\}$, where ${\bf r}$ runs over the Bravais lattice of unit cells of the honeycomb network and $a$ runs over the two sublattices ($A$ and $B$) in the unit cell shown in Fig.~\ref{Fig:honeycomb}. The GKM can be expressed in momentum space as 
\begin{eqnarray}
H_{G} &=& \sum_{{\bf k}\in B.Z.} \Psi^\dagger_{\bf k}\cdot h_{G} \cdot \Psi_{\bf k} \\
\nonumber &=&  \sum_{{\bf k} \in B.Z.} \Psi^\dagger_{\bf k} \bigg{[} \begin{pmatrix}
0 & -t f({\bf k}) - t_3 f_3({\bf k}) \\
-t f^*({\bf k}) - t_3 f_3^*({\bf k}) & 0
\end{pmatrix}
\otimes \mathbbm{1}_{2\times 2}+ \\
&& \hspace{5cm} + \begin{pmatrix}
2\lambda_{so} g({\bf k}) & 0 \\
0 & -2\lambda_{so} g({\bf k}) 
\end{pmatrix}
\otimes \sigma_z\bigg{]} \Psi_{\bf k} \\
&\equiv& \sum_{{\bf k} \in B.Z.} \Psi^\dagger_{\bf k} \left[ \mathbbm{M}_{G}({\bf k})\otimes \mathbbm{1}_{2\times 2} + 2\lambda_{so} g({\bf k}) \tau_z \otimes \sigma_z\right] \Psi_{\bf k},
\end{eqnarray}
where $ \Psi^{T}_{\bf k} \equiv ( \Psi^{\uparrow T}_{\bf k}~\Psi^{\downarrow T}_{\bf k} )= (
c_{{\bf k} \uparrow}(A) ~ c_{{\bf k} \uparrow} (B) ~ c_{{\bf k} \downarrow} (A) ~ c_{{\bf k}\downarrow} (B)
)$. $\sigma_z$ and $\tau_z$ are Pauli matrices for spin and sublattice degrees of freedom. $g({\bf k}) \equiv -\sin ({\bf k}\cdot {\bf e}_1) + \sin({\bf k}\cdot {\bf e}_2) + \sin [{\bf k} \cdot ({\bf e}_1 - {\bf e}_2)]$, $f({\bf k}) = 1 + e^{i {\bf k} \cdot {\bf e}_1} + e^{i {\bf k} \cdot {\bf e}_2}$, $f_3({\bf k}) = e^{i {\bf k}\cdot({\bf e}_1 + {\bf e}_2)} + 2\cos[{\bf k}\cdot({\bf e}_1 - {\bf e}_2)]$. 

The B.Z. is shown in Fig.~\ref{Fig:BZ}. In the KM-type models that we consider, only a few momentum points are relevant for the low-energy descriptions. Besides the usual wavevectors ${\bf K}=- {\bf K'}= (4\pi/3,0)$ in the original KM model, the TRIM points located at ${\bf M}_{1,2} \equiv (\pm \pi, \pi/\sqrt{3})$ and ${\bf M}_3\equiv (0, 2\pi/\sqrt{3})$ also need to be considered. In GKM, when we vary the third neighbor hopping strength $t_3$, we find that the band gaps close at all TRIM points, ${\bf M}_{a=1,2,3}$\cite{hung2014} as shown in Fig.~\ref{Fig:GKM_spectrum}. At the TRIM, the diagonal elements of the Hamiltonian matrices vanish, $g({\bf M}_{a}) = 0$, and the band gaps in GKM are controlled by the off-diagonal elements, which are related to real-valued first and third neighbor hoppings. Focusing on the gap closing points (TRIM), we can write down the low-energy description of GKM,
\begin{eqnarray}\label{Eq:GKM_low}
\mathcal{H}^\sigma_G = \Delta t_G \Psi^{\sigma\dagger}_{{\bf M}_a} \tau_x \Psi^\sigma_{{\bf M}_a},
\end{eqnarray}
where we introduce $\Delta t_G = t-3t_3$.  The band gap of GKM is 
\begin{eqnarray}
 \Delta \mathcal{E}_{G} = 2 \left| \Delta t_G\right|,
\end{eqnarray}
which vanishes at $t^c_3 = \frac{1}{3}t$\cite{hung2014} . At $t_3^c$, a topological phase transition takes place, from the nontrivial QSH state with $|C_{\sigma}|= 1$, $\nu=1$,  to the trivial insulating state $|C_{\sigma}|= 2$, $\nu=0$.\cite{meng2013,hung2014} 
Note that although the $|C_{\sigma}|= 2$ state is trivial in the $Z_2$ aspect, the state belongs to the classification of a two-dimensional topological crystalline insulator with mirror Chern number $C_m=2$. Under such meaning, it is still topologically nontrivial.\cite{liujunwu}

\subsection{Dimerized Kane-Mele Model}
\begin{figure}[t]
\centering
 \subfigure{\includegraphics[width=1.5 in]{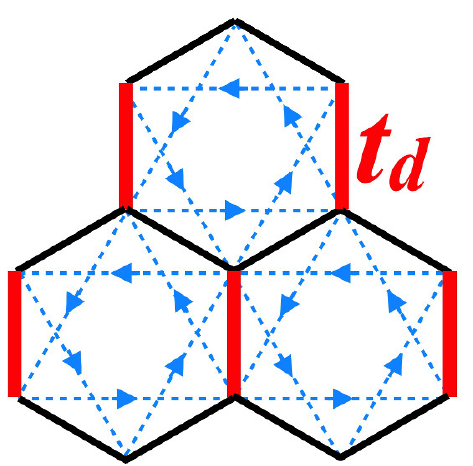} 
 \label{Fig:DKM_honeycomb}}
 \subfigure{\includegraphics[width= 1.8 in]{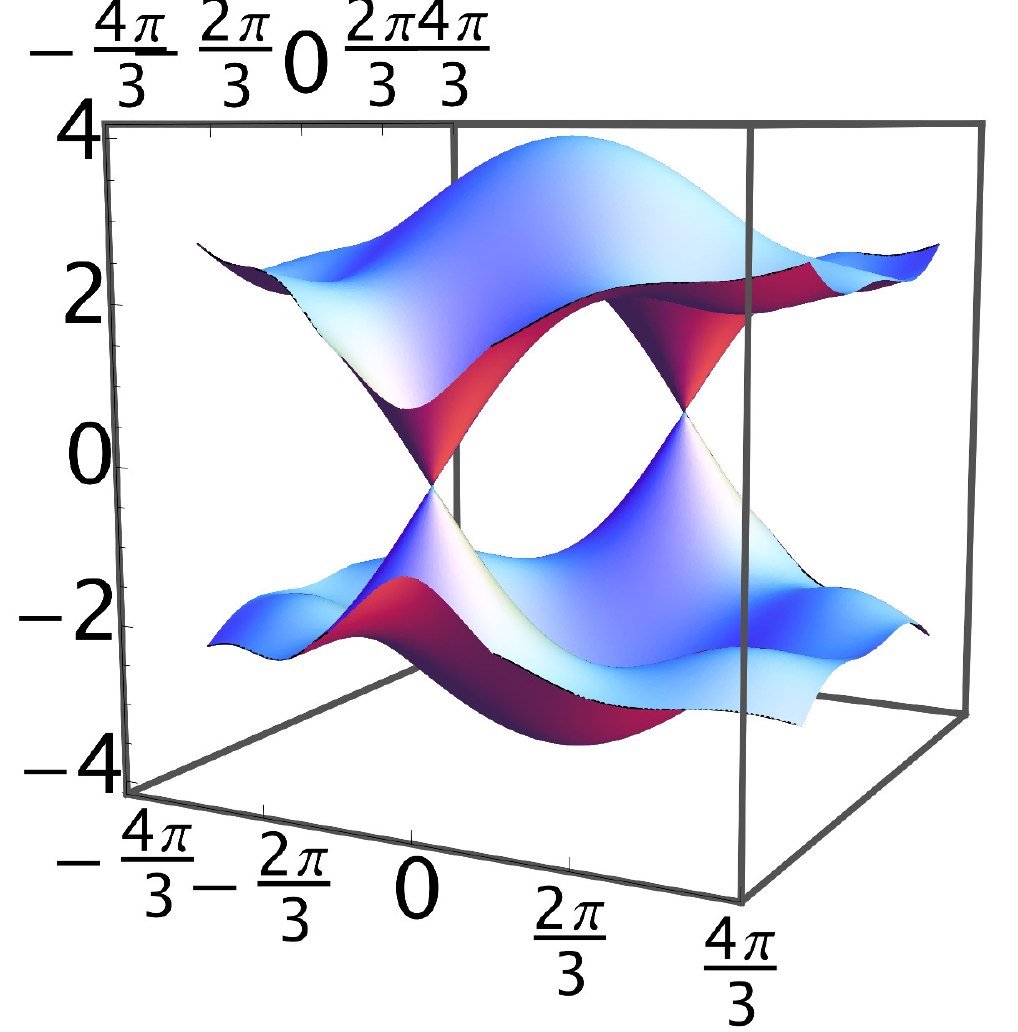}
 \label{Fig:DKM_spectrum}}
\caption{(a) Schematic of the DKM model and (b) the band spectrum at the topological phase transition. The blue dashed lines represent the imaginary second neighbor hoppings and the arrow directions represent their signs. The red lines represent the anisotropic $t_d$ hoppings with $t_d > t$, which breaks $C_3$ rotation. At the phase transition, the band gaps only close at TRIM ${\bf M}_3$ due to the breaking of $C_3$ symmetry}
  \label{Fig:DKM_model}
\end{figure}

Second variant of the KM model is the DKM. The model is given "dimerizing" one of nearest  neighbor hoppings, and the Hamiltonian $H_D$ reads as
\begin{equation}
\label{eq:H_D}
H_{D}=  -\sum_{\la jk \ra} \sum_{\sigma} t_{jk} c^\dagger_{j \sigma} c_{k \sigma} + i \lambda_{so} \sum_{\la \la jk \ra \ra } \sum_{\sigma} \sigma c^\dagger_{j \sigma} \nu_{jk} c_{k\sigma},\\
\end{equation}
where $t_{ij} = t_d~(t)$ if the two sites $\la j k \ra$ belong to the same (different) unit cell(s), and we choose $t_d~(t)>0$. The schematic of the DKM is illustrated in Fig.~\ref{Fig:DKM_honeycomb}. Similarly, we replace the site labeling $j$ with $j=\{ {\bf r},a \}$ and the DKM can be expressed in momentum space as
 \begin{eqnarray}
H_{D} &=& \sum_{{\bf k}\in B.Z.} \Psi^\dagger_{\bf k}\cdot h_{D} \cdot \Psi_{\bf k} \\
\nonumber  &=&  \sum_{{\bf k} \in B.Z.} \Psi^\dagger_{\bf k} \bigg{[} \begin{pmatrix}
0 & -t_d - t f_d({\bf k}) \\
-t_d -t f_d^*({\bf k}) & 0
\end{pmatrix} \otimes \mathbbm{1}_{2\times 2} + \\
&& \hspace{5cm} + \begin{pmatrix}
2\lambda_{so} g({\bf k}) & 0 \\
0 & -2\lambda_{so} g({\bf k}) 
\end{pmatrix}
\otimes \sigma_z\bigg{]} \Psi_{\bf k}\\
&\equiv& \sum_{{\bf k} \in B.Z.} \Psi^\dagger_{\bf k} \left[ \mathbbm{M}_{D}({\bf k})\otimes \mathbbm{1}_{2\times 2} + 2\lambda_{so} g({\bf k}) \tau_z \otimes \sigma_z\right] \Psi_{\bf k},
\end{eqnarray}
where $f_d({\bf k}) = e^{i {\bf k}\cdot {\bf e}_1} + e^{i {\bf k} \cdot {\bf e}_2}$. When we vary the dimerized hopping amplitude $t_d$ while fixing $t$, we find the band gap only closes at ${\bf M}_3$ due to the  breakdown of $C_3$ rotation as shown in Fig.~\ref{Fig:DKM_spectrum}. Similar to GKM, the band gap at TRIM (${\bf M}_3$) is also controlled by the off-diagonal terms since $g({\bf M}_3) =0$. Focusing on the ${\bf M}_3$, we can write down the low-energy descriptions of DKM,

\begin{eqnarray}\label{Eq:DKM_low}
\mathcal{H}^\sigma_D = \Delta t_D \Psi^{\sigma\dagger}_{{\bf M}_3} \tau_x \Psi^\sigma_{{\bf M}_3},
\end{eqnarray}
where $\Delta t_D = 2t-t_d$. The band gap is 
\begin{eqnarray}
\Delta \mathcal{E}_{D} = 2 \left| \Delta t_D \right],
\end{eqnarray}
which vanishes at $t^c_d = 2t$ as the critical point.  Upon increasing $t_d$, the topological phase transition turns the $C_{\sigma}=\pm 1$ QSH state  to the $C_{\sigma}=0$ trivial insulating state.\cite{meng2013,hung2014}

\subsection{Stagger-potential Kane-Mele Model}
\begin{figure}[t]
\centering
 \subfigure{\includegraphics[width=1.5 in]{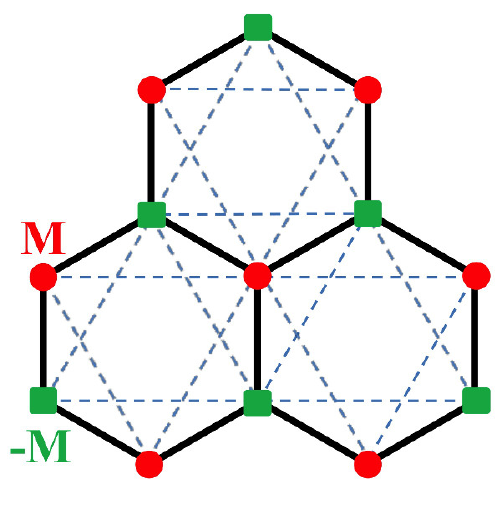} 
 \label{Fig:SKM_honeycomb}}
\subfigure{ \includegraphics[width= 1.8 in]{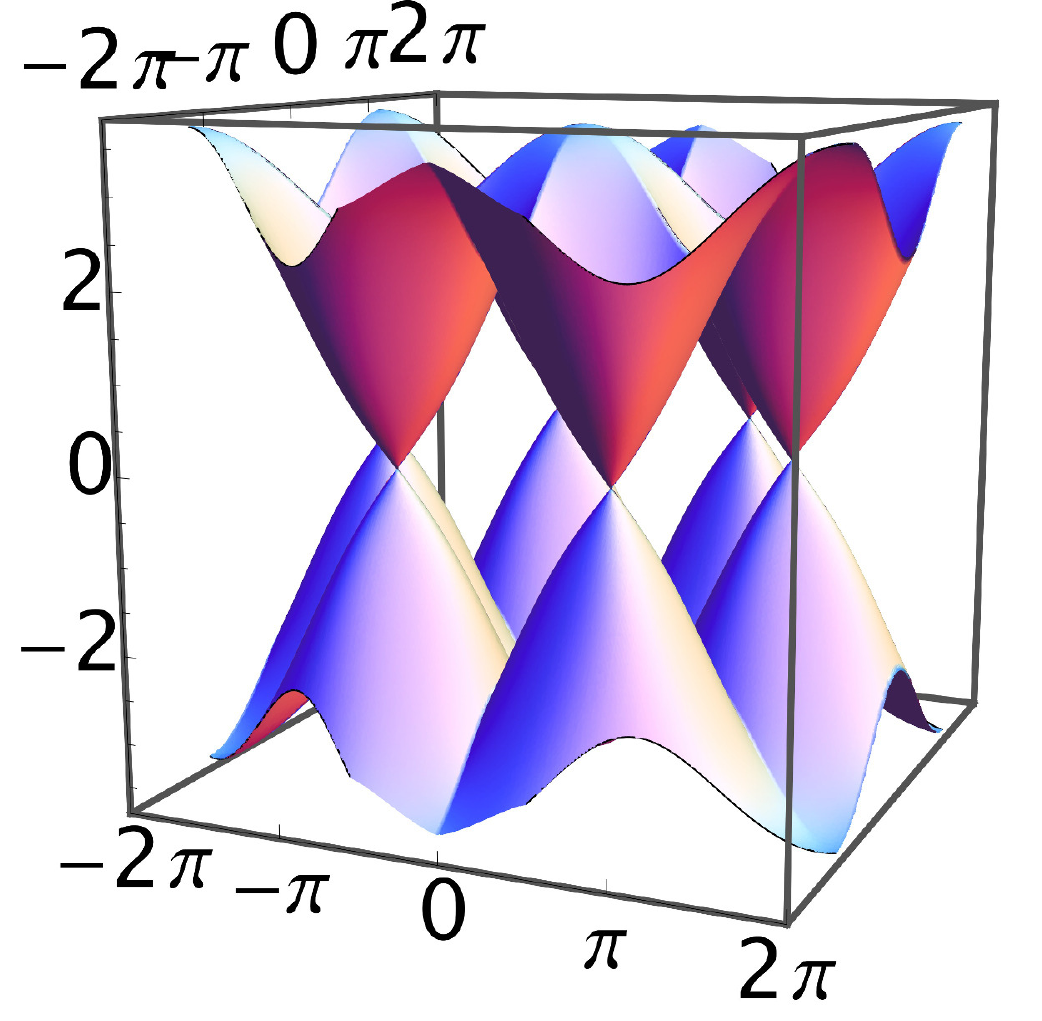}
 \label{Fig:SKM_spectrum}}
\caption{(a) Schematic of the SKM model and (b) the band spectrum at the topological phase transition. (a) The red (green) dots at sublattice $A (B)$ represent the staggered potential strength $M (-M)$. The blue dashed lines represent the imaginary second neighbor hoppings and the arrow directions represent their signs. (b) At the phase transition point, the bands gaps close at the usual locations of Dirac nodes, (${\bf K} = -{\bf K'}=(4\pi/3,0)$). The rest of the Dirac cones are related by $C_3$ rotation.}
  \label{Fig:SKM_model}
\end{figure}

So far, the GKM and DKM that we have discussed both conserve the inversion symmetry as well as PHS. Now we shift our focus on the SKM which breaks inversion symmetry due to the presence of staggered potentials. The SKM Hamiltonian, $H_S$, is
\begin{eqnarray}\label{eq:H_S}
\nonumber && H_{S}= - t \sum_{\la jk \ra} \sum_{\sigma} c^\dagger_{j \sigma} c_{k \sigma} + i \lambda_{so} \sum_{\la \la jk \ra \ra } \sum_{\sigma} \sigma c^\dagger_{j \sigma} \nu_{jk} c_{k\sigma} + M \sum_{j} \sum_{\sigma}  \epsilon_j  c^\dagger_{j \sigma} c_{j \sigma},
\end{eqnarray}
with $\epsilon_j = \pm 1$ for sublattice $j \in \{ A, B\}$.  The schematic of SKM is shown in Fig.~\ref{Fig:SKM_honeycomb}. We note that due to the presence of the staggered potentials, the PHS is explicitly broken. the SKM Hamiltonian can be expressed in the momentum space as 
\begin{eqnarray}
H_{S} &=& \sum_{{\bf k}\in B.Z.} \Psi^\dagger_{\bf k}\cdot h_{S} \cdot \Psi_{\bf k} \\
\nonumber  &=&  \sum_{{\bf k} \in B.Z.} \Psi^\dagger_{\bf k} \bigg{[} \begin{pmatrix}
M & -t f({\bf k}) \\
-t f^*({\bf k}) & -M
\end{pmatrix}
\otimes \mathbbm{1}_{2\times 2} + \\
&& \hspace{5cm} + \begin{pmatrix}
2\lambda_{so} g({\bf k}) & 0 \\
0 & -2\lambda_{so} g({\bf k}) 
\end{pmatrix}
\otimes \sigma_z\bigg{]} \Psi_{\bf k} \\
&\equiv& \sum_{{\bf k} \in B.Z.} \Psi^\dagger_{\bf k} \left[ \mathbbm{M}_{S}({\bf k})\otimes \mathbbm{1}_{2\times 2} + 2\lambda_{so} g({\bf k}) \tau_z \otimes \sigma_z\right] \Psi_{\bf k},\label{Eq:SKM_H}
\end{eqnarray}
where $f({\bf k})$ and $g({\bf k})$ are defined in GKM. While varying the strength of the staggered potentials, we find that the band gaps close at the usual locations of two independent Dirac nodes (${\bf K} = - {\bf K'} = (4\pi/3,0)$), Fig.~\ref{Fig:SKM_spectrum}.  Focusing on the gap closing points, we know there are only two gapless bands (2 gapless spin-$\downarrow$ bands at ${\bf K} $ and 2 gapless spin-$\uparrow$ bands at ${\bf K'}$). Since these two bands are related by the TRS, $\mathcal{T} : \Psi^\sigma_{\bf k} \rightarrow \epsilon^{\sigma \bar{\sigma}} \Psi^{\bar{\sigma}}_{-{\bf k}}$ with $\sigma = \uparrow(\downarrow) = 1~(2)$, we can simply focus on one spin species of fermions at one Dirac node, say ${\bf K}$. Focusing on ${\bf K}$, we can write down the effective low-energy description as
\begin{eqnarray}\label{Eq:SKM_low}
\mathcal{H}^{\downarrow}_S ({\bf K}) = \Psi_{\bf K}^{\downarrow \dagger} \begin{pmatrix}
M - 2 \lambda_{so} ({\bf K}) & 0 \\
0 & - M + 2\lambda_{so} ({\bf K})
\end{pmatrix}
\Psi_{\bf K}^{\downarrow}.
\end{eqnarray}
Therefore, the band gap of SKM is
\begin{eqnarray}
\Delta \mathcal{E}_S = 2 | M - 2 \lambda_{so}({\bf K})|.
\end{eqnarray}
We know $\lambda_{so}(\pm{\bf K})=\frac{3\sqrt{3}}{2}t$ and the gap closes at $M^c = \sqrt{3} \lambda_{so}$ in the noninteracting limit.\cite{Lai_staggerKM} Similar to the DKM variant, the topological phase transition turns the QSH state to a topologically trivial state with $|C_{\sigma}|=0$.

\section{$U/t$ expansion and mean-field decouplings}\label{Sec:PMF}
Next we start the discussions of correlation effects on the QSH state. 
The simplest correlation effect is considered by studying the variants of KMH Hamiltonian $H=H_{G,D,S}+H_U$, where $H_U$ represents the short-ranged Hubbard interaction, Eq.~\eqref{eq:H_U}. According to the low-energy descriptions for the three variants of KM models, Eqs.~(\ref{Eq:GKM_low}) and (\ref{Eq:DKM_low}) for GKM and DKM, and Eq.~(\ref{Eq:SKM_low}) for SKM, first we notice that  the gaps for GKM and DKM vanish at the TRIM, unlike the usual KM model, and are controlled by the {\it off-diagonal} elements describing the hoppings between different sublattices. On the other hand, the gaps for SKM vanish at the usual momenta ${\bf K}$ and ${\bf K'}$, and are controlled by the {\it diagonal} elements describing the more ordinary mass terms consisting of staggered potentials and the second-order hoppings (SOC). As far as the mechanism for the topological phase boundary shifts due to the presence of the short-range Hubbard interaction is concerned, we should distinguish the SKM from GKM and DKM. From the symmetry perspective, GKM and DKM both satisfy PHS while SKM breaks PHS. We will discuss these two different cases separately in the following two subsections.
\subsection{Perturbative mean-field decoupling for GKM and DKM}\label{Sec:QMC}

In the GKM and DKM, the feature that the band gaps are controlled by the off-diagonal terms describing the hoppings hints that we need to perform the perturbation up to second order. Performing the expansion in $U/t$ up to second order, we obtain the contributions to the bare Hamiltonians as $\delta H = \delta \mathcal{H}_1 + \delta\mathcal{H}_2$, where the $\delta \mathcal{H}_{1(2)}$ represent the first (second) order corrections. The $\delta \mathcal{H}_1$ term is just the first-order mean-field decouplings given by Eq. (\ref{Eq: Hubbard_1st_order}).
 As suggested by the QMC results,\cite{Hung2013,hung2014} no charge density wave orders and magnetic instability are found at intermediate interaction strengths. Thus, we preserve the translational symmetry, and set $\la s_z({\bf r},a)\ra=0$ as well as $\la n({\bf r},a) \ra= \la n (a)\ra \equiv \la n_a \ra$. In momentum space,  $\delta \mathcal{H}_1 ({\bf k}) = \sum_{{\bf k} \in B.Z.} \Psi^\dagger_{\bf k} h_1({\bf k}) \Psi_{\bf k}$, with 
\begin{eqnarray}\label{Eq:Hubbard_1st_order_h1}
h_1 ({\bf k}) = \frac{U}{2}\begin{pmatrix}
 \la n_A \ra \\
 0 &  \la n_B \ra
 \end{pmatrix}
 \otimes \mathbbm{1}_{2\times2} = \frac{U\la n \ra}{2}  \mathbbm{1}_{4 \times 4},
 \end{eqnarray}
 where we explicitly used the fact that $\la n_A \ra = \la n_B \ra = \la n \ra$ above.
 
The second-order correction $\delta\mathcal{H}_2$ consists of two terms, $\delta\mathcal{H}_2 = \delta H^{(1)}_2 + \delta H^{(2)}_2$:
\begin{eqnarray}
\delta H^{(1)}_2 = -\frac{U^2}{2}\sum_{{\bf r}, {\bf r}', a}n_{\uparrow}({\bf r},a) n_{\uparrow}({\bf r}', a) n_{\downarrow}({\bf r},a)n_{\downarrow}({\bf r}',a),
\end{eqnarray}
and
\begin{eqnarray}
\delta H^{(2)}_2 = -U^2 \sum_{{\bf r}, {\bf r}'} n_\uparrow ({\bf r}, A) n_\uparrow ({\bf r}',B) n_\downarrow ( {\bf r}, A) n_\downarrow ({\bf r}',B).
\end{eqnarray} 
For simplicity in performing PMF for $\delta H^{(1)}_2$, we assume ${\bf r'} = {\bf r} + \vec{E}_{\mu}$, where $\vec{E}_\mu$ runs over the Bravais lattice of the unit cell that is connected to ${\bf r}$. The PMF gives
\begin{eqnarray}
\nonumber \delta H^{(1)}_2 =  && \frac{U^2}{2}  \sum_{{\bf k}, \vec{E}_\mu,\sigma, a}  \bigg{[} \bigg{(} \la n_\sigma(a)\ra \left| \chi_{\ell \bar{\sigma}}(\vec{E}_\mu, a ) \right|^2 - e^{-i {\bf k}\cdot \vec{E}_\mu} \chi_{\ell\sigma}(\vec{E}_\mu, a)  \times \\
&&\hspace{4cm} \times \left| \chi_{\ell \bar{\sigma}} (\vec{E}_\mu , a ) \right|^2 \bigg{)}  c^\dagger_{{\bf k} \sigma} (a) c_{{\bf k} \sigma}(a ) + \Hc\bigg{]},\label{Eq:H2-1}
\end{eqnarray}
where we define the $\ell$-neighbor hopping correlation $[\chi_{\ell \sigma}(\vec{E}_\mu,a)]^* \equiv \la c^\dagger_\sigma ({\bf r} + \vec{E}_\mu, a) c_\sigma ( {\bf r},a)\ra$, with $\ell$ being the number of sites covered by $\vec{E}_\mu$. For correlations between the same sublattices, $\ell$ is always even. $\bar{\uparrow} = \downarrow$ and vice versa. For the GKM and DKM models, we restrict $\ell=2$ for second neighbor hopping (SOC) renormalization and $\vec{E}_\mu = \{{\bf e}_1,~{\bf e}_2,~{\bf e}_3 \equiv {\bf e}_1 - {\bf e}_2\}$, which is enough to capture the essential physics of the QMC results.  Under PMF, $\delta H^{(1)}_2$ only renormalizes the diagonal terms of the Hamiltonian matrices.

For the PMF of the $\delta H^{(2)}_2$, we introduce $({\bf r}',a) = ({\bf r}+ \vec{E}_\nu,a)$, with $\vec{E}_\nu$ being the vectors connected to ${\bf r}$. Note that $\vec{E}_\nu$ contain ${\bf e}_0 \equiv {\bf 0}$, which means the two sites are in the same unit cell. We obtain
\begin{eqnarray}
\delta H^{(2)}_2 =U^2 \sum_{{\bf k}, \vec{E}_\nu, \sigma} \bigg{[}  e^{-i {\bf k} \cdot \vec{E}_\nu}  \chi_{m \sigma} (\vec{E}_\nu, AB) \left| \chi_{m \bar{\sigma}}(\vec{E}_\nu,AB) \right|
^2 c^\dagger_{{\bf k} \sigma} (B) c_{{\bf k} \sigma} (A) + \Hc \bigg{]},~~~~\label{Eq:H2-2}
\end{eqnarray}
where $[\chi_{m \sigma} (\vec{E}_\nu, AB)]^* \equiv \la c^\dagger_\sigma ({\bf r} + \vec{E}_\nu , B) c_\sigma ({\bf r},A)\ra$, with $m$ being the number of sites covered by $\vec{E}_\nu$. Since $\vec{E}_\nu$ connects two different sublattices, $m$ is always odd. For simplicity, we restrict $m=1,~3$ for the GKM to capture the renormalizations of the first and third neighbor hoppings and $m=1$ for the DKM. For more efficient numerical calculations, we can utilize symmetries [$C_2$, Inversion + complex conjugation ($\mathcal{I}^*$), TRS, PHS for both GKM and DKM while there is an additional $C_3$ for GKM] to reduce the number of variables in each model. 

Above in Eqs. (\ref{Eq:Hubbard_1st_order_h1}), (\ref{Eq:H2-1}), and (\ref{Eq:H2-2}), we performed the perturbation up to $O(U^2/t^2)$ and utilized symmetry arguments to simplify the expressions of the corrections to the bare Hamiltonians as functions of hopping correlations $\chi-s$. For determining the shift of the phase transition location, we will rely on the low-energy descriptions around the gap closing points, located at TRIM for GKM/DKM and $\pm{\bf K}$ for SKM. Focusing on the gap closing points, we will define the {\it renormalized} gap equations for three variants of KM models below and later we will numerically solve for the parameters $\chi$-s to determine how the bands gaps get renormalized.

\subsubsection{PMF for the GKM model and the gap equation}
 For the hoppings between different sublattices, we choose $\vec{E}_\nu = \{ {\bf e}_0 ,~{\bf e}_1, ~{\bf e}_2\}$ for $m=1$ and $\vec{E}_\nu = \{ \pm ({\bf e}_1 - {\bf e}_2),~{\bf e}_1 + {\bf e}_2\}$ for $m=3$. We can simplify Eqs. (\ref{Eq:H2-1})-(\ref{Eq:H2-2}) by identifying $\chi_{m \sigma}( \vec{E}_\nu ,AB) = \chi_{m \sigma}= \chi_m$, $\chi_{2\sigma}({\bf e}_1,a) = - \chi_{2\sigma}({\bf e}_2,a) = - \chi_{2\sigma} ({\bf e}_1 - {\bf e}_2,a)$, and $\chi_{2\sigma} ({\bf e}_\mu ,a ) = \chi^*_{2\bar{\sigma}}({\bf e}_\mu, a) = - \chi_{2\bar{\sigma}}({\bf e}_\mu,a)$, where we use the fact that $\chi_{2\sigma}({\bf e}_\mu,a) \in \mathbbm{I}$. For clarity, we introduce $\chi^{*}_{2\uparrow}({\bf e}_1, a ) = i \chi_{2\uparrow}(a)$ and $\chi_{2\sigma}(A) = - \chi_{2\sigma}(B)\equiv \chi_2 \in \mathbbm{R}$. 
\begin{eqnarray}
&& \delta H^{(1)}_2 = U^2 \sum_{{\bf k},\sigma,a } \bigg{[} 3  \la n_\sigma (a)\ra   \chi_2 ^2  - (-1)^{\sigma + a }  g({\bf k}) \chi_2^3\bigg{]} c^\dagger_\sigma ({\bf k},a) c_\sigma({\bf k},a). \label{Eq:GKM_2nd_1}\\
&& \delta H^{(2)}_2 = - U^2 \sum_{{\bf k}, \sigma} \bigg{[} f({\bf k})  \chi_1^3  + f_3({\bf k}) \chi_3^3 \bigg{]} c^\dagger_\sigma ({\bf k}, A ) c_\sigma ({\bf k}, B ) + \Hc. \label{Eq:GKM_2nd_2}
\end{eqnarray}
At $\pm{\bf K}$, $g(\pm {\bf K})$ vanishes. Eq.~(\ref{Eq:GKM_2nd_1}) simply renormalizes the chemical potential and Eq.~(\ref{Eq:GKM_2nd_2}) renormalizes the off-diagonal elements (renormalize the hopping amplitudes).

After we add the corrections due to the interaction into the bare Hamiltonian, we focus on the low-energy descriptions near the gap closing momenta TRIM: ${\bf M}_{a=1,2,3}$. At ${\bf M}_a$, we find $g({\bf M}_a) =0$, $f({\bf M}_a) = -1$, and $f_3({\bf M}_a) = 3$. We then define the gap equation near the TRIM as
\begin{eqnarray}\label{Eq:gap_GKM}
\Delta_{G} = t - 3 t_3 + U^2\bigg{(} \chi_1^3  - 3 \chi_3^3  \bigg{)}.
\end{eqnarray}
For the noninteracting critical point, $t_3 = \frac{1}{3} t$.  At weak-coupling, $U/t \ll 1$, we can approximate $\chi_1$ and $\chi_3$ to be the noninteracting values. We find that $\chi_1 \simeq 0.20705$ and $\chi_3 \simeq 0.03064$ and the $U^2$ correction is roughly $0.00879 U^2$. We conclude that in the weak-coupling limit, the topological phase is more stable against the third neighbor hoppings since we need larger $t_3$ to close the gap, coinciding with the QMC result.\cite{Hung2013,hung2014}

\subsubsection{PMF for the DKM model and the gap equation}
For the hoppings between different sublattices we only need to consider the renormalizations of the first neighbor hoppings $t$ and $t_d$ with $m=1$.  Since the $C_3$ rotation is broken, the hopping amplitudes within a unit cell are no longer equivalent to those between different unit cells. Utilizing symmetry considerations, we can identify the hopping amplitudes within the same unit cell $\chi^\uparrow_1({\bf e}_0,AB)=\chi^\downarrow_1({\bf e}_0,AB) \equiv \chi^d_1$. For the hopping between different unit cells $\chi^\sigma_1({\bf e}_1,AB) = \chi^\sigma_1({\bf e}_2,AB)\equiv \chi_1$. For the second neighbor hopping, we have $\chi_{2\sigma} ({\bf e}_1, a) = - \chi_{2\sigma}({\bf e}_2,a) \not = \chi_{2\sigma}({\bf e}_3,a)$, $ \chi_{2\uparrow}({\bf e}_{\mu=1,2,3}, a) = -\chi_{2\downarrow}({\bf e}_\mu,a)$, and $\chi_{2\sigma}({\bf e}_\mu, A) = - \chi_{2\sigma}({\bf e}_\mu, B)$. For clarity, we define $\chi^*_{2 \uparrow}({\bf e}_1, a)=-\chi^*_{2\uparrow}({\bf e}_2,a) \equiv i \chi_{2\uparrow} (a)$, $\chi^*_{2\uparrow} ({\bf e}_3, a) \equiv i \chi^d_{2\uparrow}(a)$. We further introduce $\chi_{2\uparrow}(a) \equiv (-1)^{a+1} \chi_2 $ and $\chi^d_{2\uparrow}(a) \equiv (-1)^{a+1} \chi^d_2$, with $a=A~(B) = 1~(2)$. The second-order corrections to the bare Hamiltonian are
\begin{eqnarray}
\nonumber && \delta H^{(1)}_2  = U^2 \sum_{{\bf k}, \sigma, a} \bigg{\{}  \la n_\sigma (a)\ra \bigg{(}  2 \chi_2^2 + (\chi^d_2)^2 \bigg{)}  - (-1)^{\sigma + a } \bigg{[}\bigg{(} -\sin({\bf k}\cdot {\bf e}_1) + \sin({\bf k}\cdot {\bf e}_2)\bigg{)} \times \\
&& \hspace{5cm} \times \chi_2^3 + \sin({\bf k}\cdot({\bf e}_1 - {\bf e}_2 ))(\chi^d_2)^3 \bigg{]}\bigg{\}} c^\dagger_\sigma ({\bf k},a) c_\sigma ({\bf k},a) \\
 && \delta H^{(2)}_2 = -U^2 \sum_{{\bf k},\sigma}  \bigg{[} (\chi^d_1)^3  + \big{(} e^{i{\bf k} \cdot {\bf e}_1} + e^{i {\bf k} \cdot {\bf e}_2}\big{)} \chi_1^3  \bigg{]} c^\dagger_\sigma({\bf k}, A) c_\sigma ({\bf k},B) + \Hc,
 \end{eqnarray}
where $\sigma = \uparrow (\downarrow) = 1 (2)$.

For the DKM model, the gap closes at ${\bf M}_3$ only. We focus on the gap closing momentum ${\bf M}_3$ and we can define the corresponding gap equation as
\begin{eqnarray}\label{Eq:gap_DKM}
\Delta_{D} = 2t - t_d - U^2 \bigg{[} (\chi^d_1 )^3 - 2 \chi_1 ^3 \bigg{]}.
\end{eqnarray}
Focusing on the critical point, $t_d = 2 t$, at the $U/t \ll 1$, we find that $\chi_1 \simeq 0.15770$ and $\chi^d_1 \simeq 0.36627$. The $U^2$ correction to the gap equation is $ -0.04129 U^2 < 0$. We conclude that in the weak-coupling regime the topological phase is more fragile to the dimerization. This is also consistent with the observation from the QMC study.\cite{lang2013,hung2014}

\subsection{Perturbative mean-field decoupling for SKM}
For SKM, since the bands gaps are controlled by the usual diagonal elements of the Hamiltonian, it is sufficient to perform the perturbation to first-order $(O(U/t))$ and then apply usual mean-field decoupling, which is already given in Eqs.~(\ref{Eq: Hubbard_1st_order}) and (\ref{Eq:Hubbard_1st_order_h1}). We find the $O(U/t)$ term can give corrections to the diagonal terms leading to a shift of the $Z_2$ topological phase transition in the SKM.

In SKM, we focus at the gap closing momenta $\pm{\bf K}$. Focusing on ${\bf K}$, we know that the off-diagonal elements in the Hamiltonian, Eq.~(\ref{Eq:SKM_H}), become linear in ${\bf k}$ and vanish exactly at ${\bf K}$. We find that two of the four bands with eigenvalues $E_1 = M -2 \lambda_{so} g({\bf K}) + U\la n_A \ra/2$ and $E_2 = - M + 2 \lambda_{so} g({\bf K}) + U \la n_B \ra/2$ get inverted by tuning the mass $m$ and $\lambda_{so}$. We can define the gap function $\Delta_S ({\bf K}) \equiv \Delta_S$ as
\begin{eqnarray}\label{Eq:gap_SKM}
\Delta_{S}({\bf K}) = 2M - 4 \lambda_{so}g({\bf K}) - \frac{U}{2}\bigg{(} \la n_B \ra - \la n_A \ra \bigg{)}.
\end{eqnarray}
For a constant staggered potential $M$ and SOC $\lambda_{SO}$, the sign of the $U$ correction term is determined by the sign of $U$ (with $\la n_B \ra > \la n_A \ra$ assumed).
 As a consequence, a repulsive interaction will stabilize the QSH phase, whereas an attractive interaction destabilizes in the SKM model. This feature is significantly different from the GKM and DKM models, where the gap equations have corrections proportional to $U^2$, cf Eqs. (\ref{Eq:gap_GKM}) and (\ref{Eq:gap_DKM}). In addition, unlike the case here, due to the PHS in the half-filled GKM and DKM models, the topological phase boundary shifts in GKM and DKM are independent of the sign of interactions.

\subsection{Self-consistent perturbative mean-field calculations}\label{Subsec:MF_numerics}
\begin{figure}[t]
\centering
   \subfigure{\includegraphics[width=2.4 in]{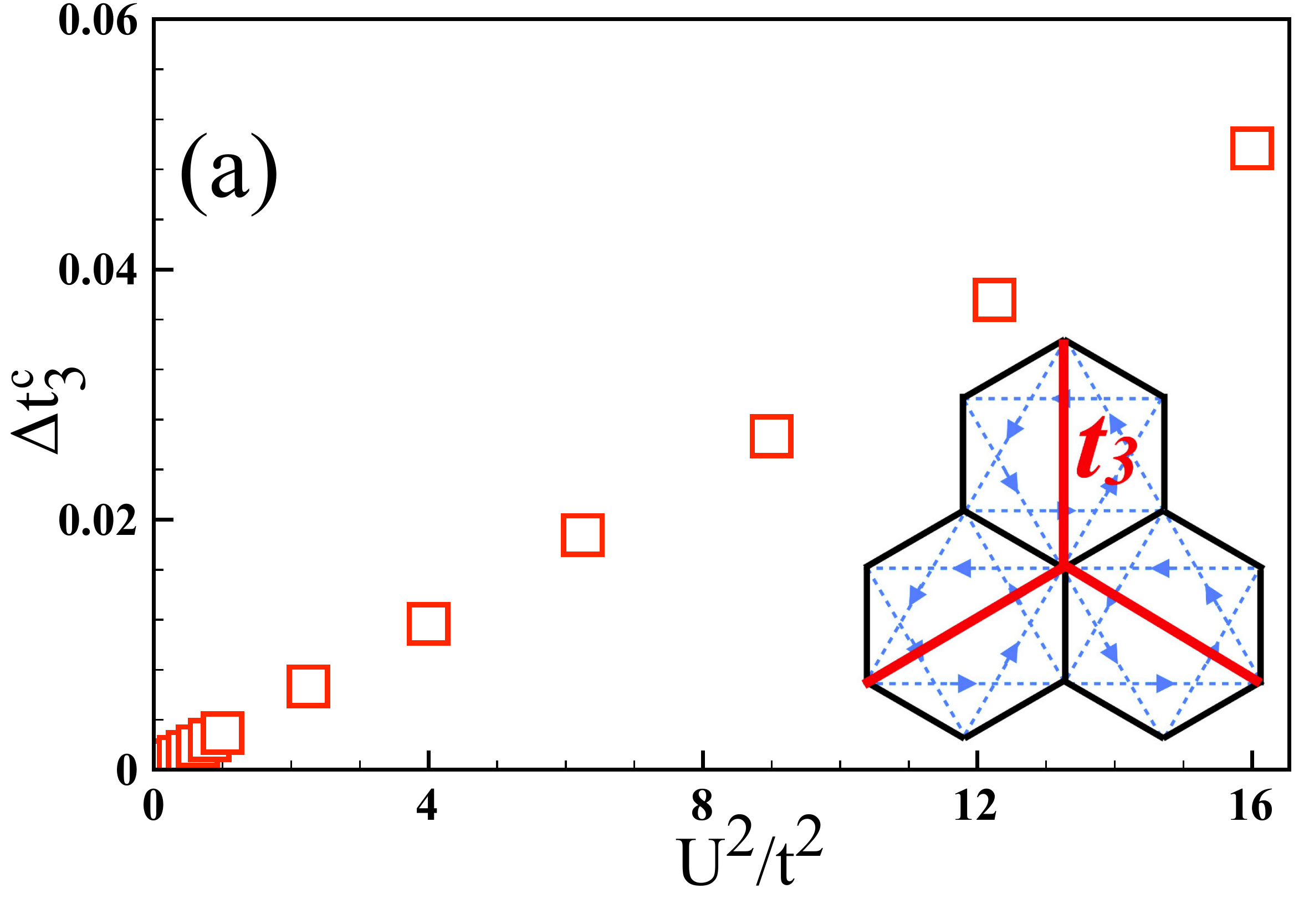}
      \label{Fig:GKM}}
   \subfigure{\includegraphics[width=2.4 in]{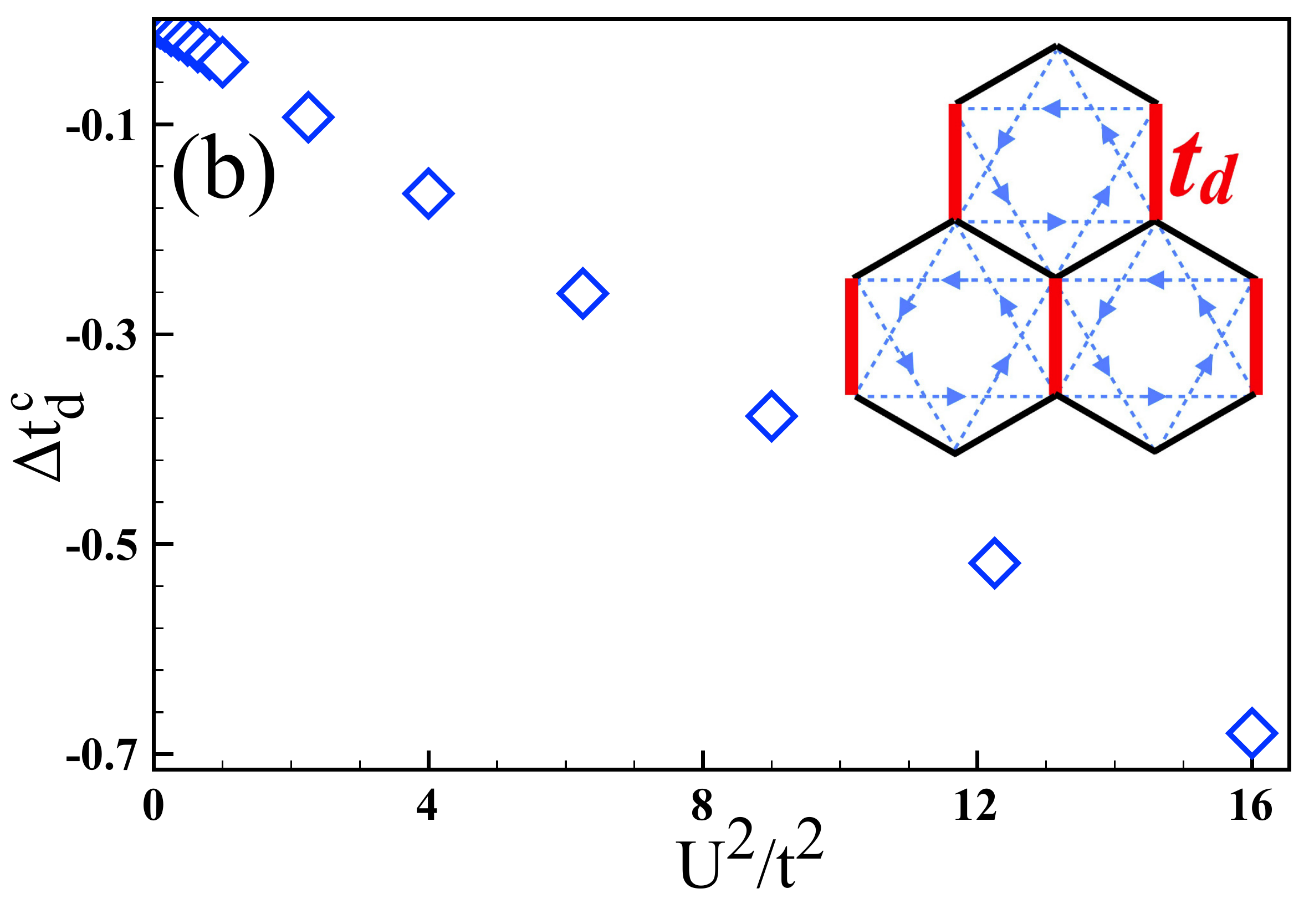} 
     \label{Fig:DKM}}
\caption{Self-consistent PMF  data for QSH boundary shifts for (a) $\Delta t^c_3=t^c_3(U,\chi)-1/3t$ in the GKM model and (b) $\Delta t^c_d=t^c_d(U,\chi)-2t$ in the DKM model, where $t^c_3(U,\chi), t^c_d(U,\chi)$ are the renormalized hopping amplitudes which are functions of $U$ and the two-point correlation functions $\chi_{ij}$ between sites $i$ and $j$. Both cases show that the amount of the boundary shifts are linearly proportional to $U^2/t^2$. In (a), the inset is the illustration of GKM model on the honeycomb lattice. The red lines represent the $t_3$ hoppings. In (b), the inset represent the DKM model on the honeycomb lattice with anisotropic hoppings that breaks $C_3$ rotation, and the red lines represent the $t_d$ hoppings with $t_d > t$. A positive shift (open red squares) indicates that the topological phase is stabilized; a negative shift (open blue diamonds) it is destabilized.}
\label{Fig:GKMDKM}
\end{figure}

In the self-consistent numerical calculations, we set the honeycomb lattice consist of $400 \times 400$ unit cells and set $t=1$ and $\lambda_{so} = 0.4$. The results at finite $U/t >0 $ for the GKM and DKM are shown in Figs.~\ref{Fig:GKM}-\ref{Fig:DKM}. The x-axis is the square of the interaction strength and the y-axis is the boundary shift amount, $\Delta t^c_3=t^c_3(U,\chi)-t^c_3(0,\chi)$ and $\Delta t^c_d=t^c_d(U,\chi)-t^c_d(0,\chi)$, where we introduced the renormalized $t^c_3(U,\chi)$ and $t^c_d (U,\chi)$ in the presence of the interaction and $\chi$ are self-consistently determined in each case. $t^c_3(0,\chi)=\frac{1}{3}t$ and $t^c_d(0,\chi)=2t$ are respectvely the topological phase boundaries for the GKM and DKM in the noninteracting limit. Since half-filled GKM and DKM models {\it preserve} PHS, QMC can have positive-definitive sampling. The results can be directly compared those obtained in the QMC presented in the latter sections.\cite{lai2014}

The positive slope in Fig.~\ref{Fig:GKM} indicates that upon increasing $U$, the critical value of $t^c_3$ moves towards a larger value ($>t^c_3(0,\chi)$).  Thus it suggests  that the onsite interaction stabilizes the QSH against the third neighbor hopping $t_3$ for GKM, consistent with our previous weak-coupling picture. On the other side, Fig~\ref{Fig:DKM} shows that the interaction makes the QSH more fragile to the dimerization $t_d$ due to the negative slope of the $\Delta t^c_d-U^2$ curve\cite{hung2014}. In addition, within PMF, we find that the hopping amplitudes are almost independent of $U/t$ and, hence, the amount of boundary shift is linearly proportional to the $(U/t)^2$.\cite{lai2014}

\begin{figure}
\centering
   \subfigure{\includegraphics[width=2.4 in]{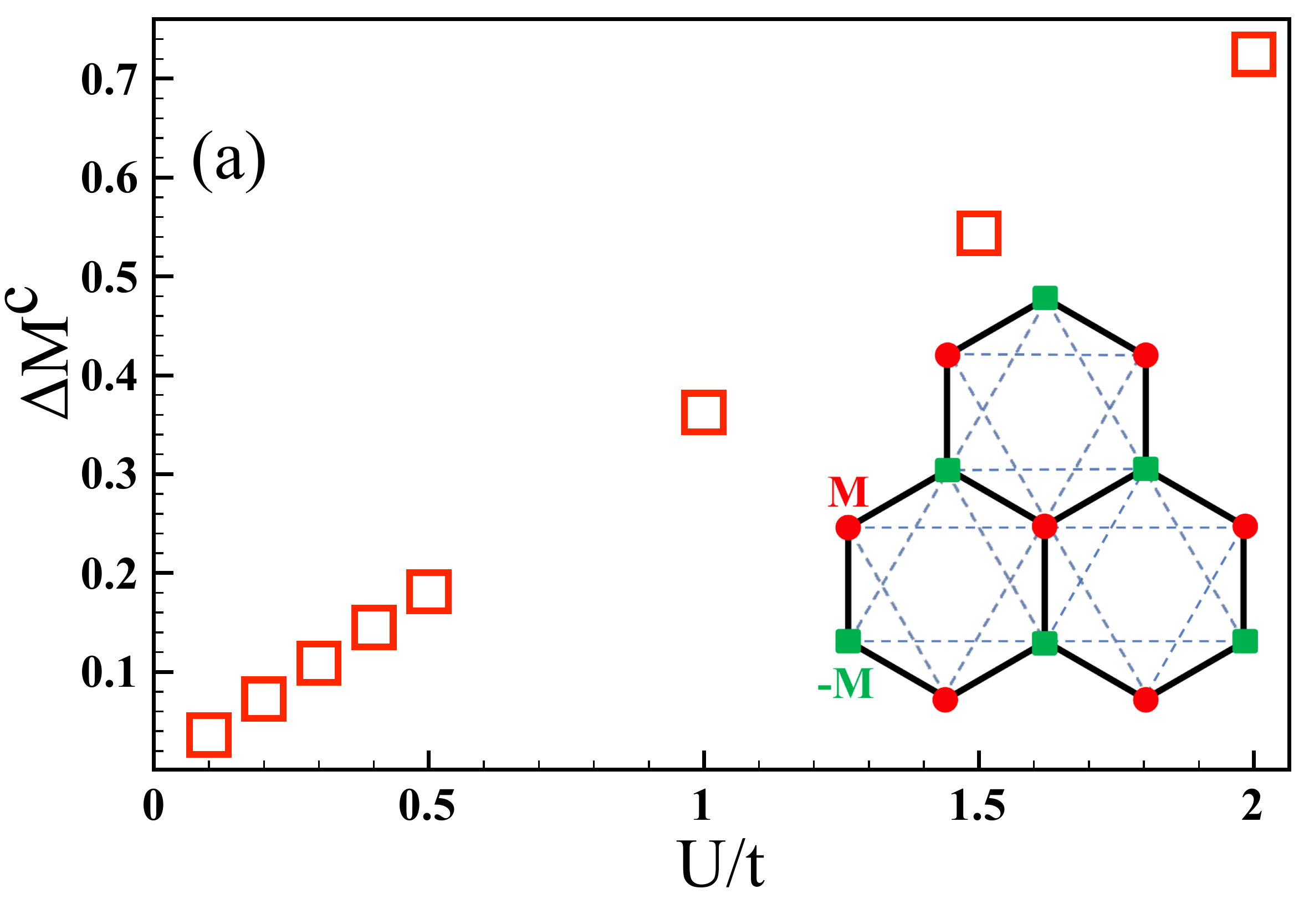}
      \label{Fig:ShiftSKM_pos}}
   \subfigure{\includegraphics[width=2.4 in]{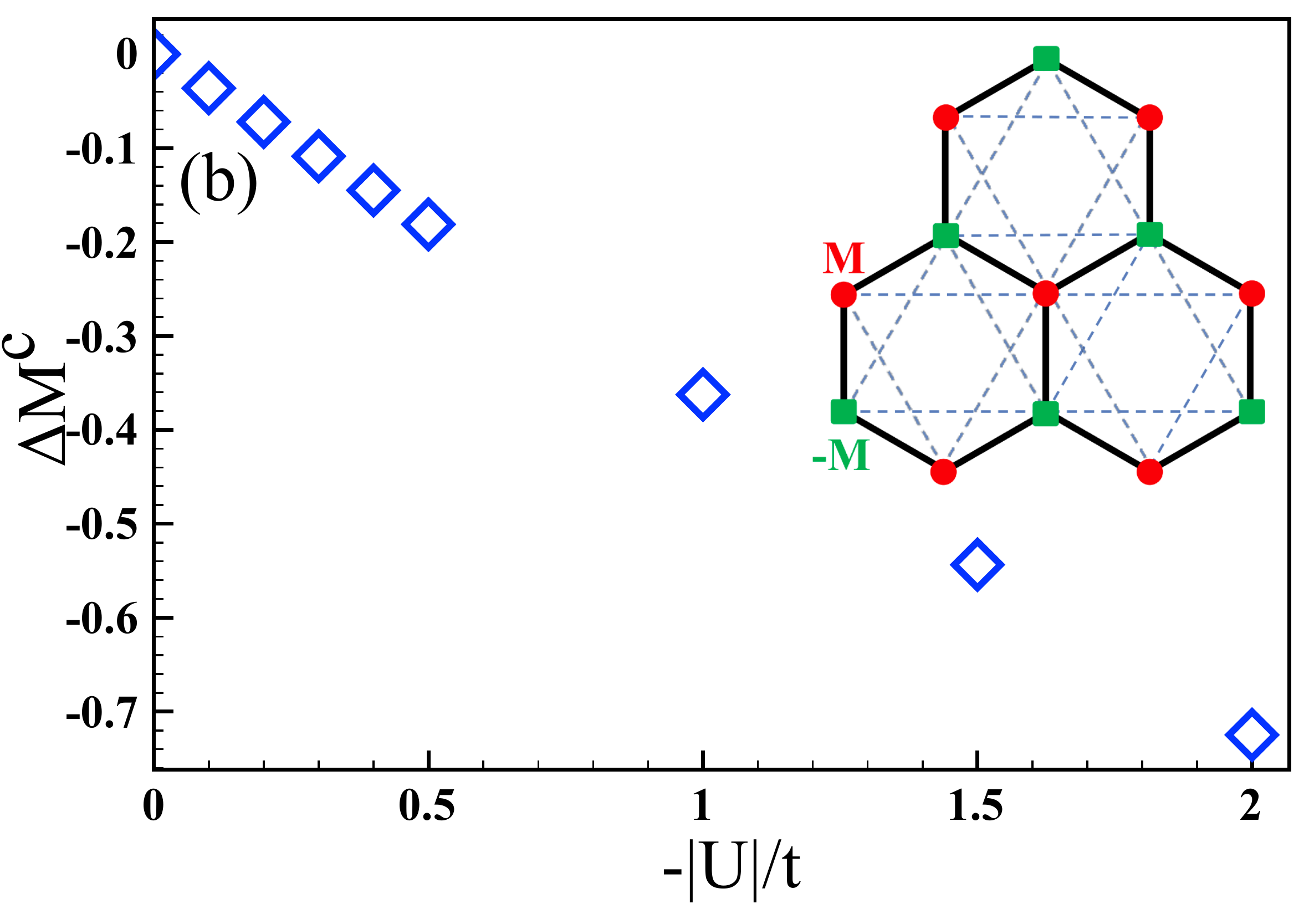} 
     \label{Fig:ShiftSKM_neg}}
\caption{Self-consistent PMF data for QSH boundary shift $\Delta M^c=M^c(U)-\sqrt{3}\lambda_{SO}$ in the SKM models with (a) $U/t >0$ and (b) $U/t =-|U|/t<0$. Note that in both cases, the amounts of the boundary shift, red open squares, are linearly proportional to $U/t$. }
\label{Fig:ShiftSKM}
\end{figure}

On the other hand, the SKM model breaks PHS and the results at positive $U/t>0$ can not be directly compared with those obtained in QMC since QMC has the sign problem. We therefore consider the both positive $U/t >0$ and negative $U/t<0$ cases in the SKM model. The self-consistent PMF results in $U/t <0$ case can be compared with the results in QMC, while the results for $U/t>0$, though can not directly be compared with QMC results, are presented for completeness. The results at finite $U/t$ are shown in Fig.~\ref{Fig:ShiftSKM}. For $U/t >0$, Fig.~\ref{Fig:ShiftSKM_pos} shows that under PMF the topological phase is more stabilized against the staggered potential strength and the shift of the topological phase boundary is linearly proportional to $U/t$. On the other hand, for $U/t = - |U|/t <0$, the topological phase is more fragile to the staggered potential strength.

\section{Sign-free determinant projector QMC}\label{Sec:QMC}
To further verify the PMF theory, we perform sign-free projective QMC simulations\cite{white1989,sorella1989,meng2010} for the variants of the KM models with onsite interactions, and then compare with the PMF results.
In the projector algorithm,  an observable $O$ is measured as
\begin{eqnarray}
\langle O \rangle =\lim_{\Theta \to \infty}\frac{\langle \Psi_\text{T} |e^{- \frac{\Theta}{2}H} O e^{- \frac{\Theta}{2}H}|\Psi_\text{T}\rangle}{\langle \Psi_\text{T} | e^{-\Theta H}| \Psi_\text{T}\rangle},
\label{eqn:expection}
\end{eqnarray}
where $|\Psi_\text{T}\rangle$ is a trial wave function. The ground state wave function $|\Psi_0\rangle$ are filtered out through applying the projection operator $e^{-\frac{\Theta}{2}H}$ onto $| \Psi_{\text{T}}\rangle$; thus we require $\langle \Psi_\textrm{T}|\Psi_0\rangle \ne 0$. For the variants of the KM model, the lowest single-particle states of $H_G$, $H_D$ and $H_S$ are good candidates for $|\Psi_\text{T}\rangle$ in the simulations. 
$\Theta$ is the projective parameter and plays the role as the imaginary time axes in the QMC algorithm. In practice, we discretize $\Theta$ into $M$ tiny slices with $\Delta \tau$ to rewrite  the projection operator as $e^{-\Theta H}=[e^{-\Delta \tau H}]^M$, where  $\Theta=\Delta\tau M$ and $\Delta\tau \ll 1$.  Further by the first-order Suzuki-Trotter decomposition, $e^{-\Delta \tau H}$ can be decomposed as
\begin{eqnarray}
e^{-\Delta \tau H} \simeq e^{-\Delta \tau H_\text{G,D,S} }e^{-\Delta \tau H_U}.
\label{eqn:suzuki}
\end{eqnarray}

The first term can be expressed as a matrix in terms of spin-particle basis, but, however, the second term $e^{-\Delta \tau H_U}$ involves non-bilinear fermionic operators, i.e. $H_U\sim c^{\dag}_{\uparrow}c_{\uparrow}c^{\dag}_{\downarrow}c_{\downarrow}$. We need to resort to discrete Hubbard-Stratonovich (HS) transformations,\cite{hirsch1983,assaad1998,assaad2002}  to transform  $e^{-\Delta \tau H_U}$ into a bilinear form by introducing auxiliary fields. For  $U>0$, we can have a two-component HS transformation\cite{hirsch1983}
\begin{eqnarray}
e^{-\Delta \tau
\frac{U}{2}(n-1)^2}=\frac{1}{2}\sum_{s=\pm1}\,e^{i\alpha s(n-1)},\label{eqn:HStransformation}
\end{eqnarray}
where $\alpha=\cos^{-1}{(e^{-\Delta \tau \frac{U}{2}})}$. We can also use the four-component counterpart\cite{assaad1998}
\begin{eqnarray}
e^{-\Delta \tau
\frac{U}{2}(n-1)^2}=\frac{1}{4}\sum_{s=\pm1,\pm2}\gamma(s)\,e^{i \sqrt{\Delta
\tau \frac{U}{2}}\,\eta(s)(n-1)}+O(\Delta
\tau^4),\label{eqn:HStransformation}
\end{eqnarray}
where
\begin{eqnarray}
   \gamma(\pm1) & = \;\; (1+\sqrt{6}/3)\;,\quad\quad\eta(\pm1) & = \;\; \pm\sqrt{2(3-\sqrt{6})}\;,\nonumber\\
   \gamma(\pm2) & = \;\; (1-\sqrt{6}/3)\;,\quad\quad\eta(\pm2) & = \;\; \pm\sqrt{2(3+\sqrt{6})}\;.
\end{eqnarray}
For the latter one, the systematic error of the HS transformation of
order  can be controlled by selecting appropriately
small values for $\Delta\tau$. 
For  $U<0$, we resort to the other two-component HS transformation\cite{hirsch1983}
\begin{eqnarray}
e^{\Delta \tau
\frac{|U|}{2}(n-1)^2}=\frac{1}{2}\sum_{s=\pm1}\,e^{\alpha^{\prime} s(n-1)},
\label{eqn:HStransformation}
\end{eqnarray}
where  $\alpha^{\prime}=\cosh^{-1}{(e^{\Delta \tau \frac{|U|}{2}})}$. In most cases, $\Delta \tau t=0.1$ and $\Delta\tau t=0.05$ are chosen in QMC simulation.

For a $N$-site system, the cost of the two-component HS transformation is to introduce $2^{NM}$ auxiliary fields.  For the four-component counterpart, the cost is about having $4^{NM}$ auxiliary fields. The integration over all auxiliary field configurations
$\lbrace s_{i,\tau} \rbrace$ is performed using stochastic Monte Carlo sampling.
In Eq. (\ref{eqn:expection}) with the two-component HS transformations, the partition function $\langle \Psi_0 | \Psi_0 \rangle$ is evaluated as
\begin{eqnarray}
   \langle\Psi_0|\Psi_0\rangle
   & = & \langle\Psi_{\text{T}}|e^{-\Theta H}|\Psi_{\text{T}}\rangle = \langle\Psi_{\text{T}}|\prod_{\tau=1}^{M} e^{-\Delta\tau H_\text{G,D,S}} e^{-\Delta\tau H_{U}}|\Psi_{\text{T}}\rangle \nonumber\\
    & = & \lim_{\Theta\to\infty}\sum_{\{s_{i,\tau}\}}\prod_{i,\tau}\; \,\prod_{\sigma}w_{\sigma}(s_{i,\tau})\;.
   \label{Trace}
\end{eqnarray}
The summation $\sum_{\{s_{i,\tau}\}}$ runs over possible auxiliary
configurations $s_{i,\tau}$, where $i = 1 \sim N$, $\tau = 1 \sim
M$.
The probability weight for spin-$\sigma$ reads as
\begin{equation}
   w_{\sigma} = \mbox{Tr}\left[
            \prod_{\tau=1}^{M}
            \exp\{-\Delta\tau\sum_{i,j}c^{\dagger}_{i,\sigma} [H^{\sigma}_{G,D,S}]_{i,j} c_{j,\sigma}\}
            \exp\{\tilde{\alpha}\sum_{i=1}^{N}
            s_{i,\tau}\Big(n_{i\sigma}-\frac{1}{2}\Big)\}
      \right],
      \label{weightsigma}
\end{equation}
with $\tilde{\alpha} = i \alpha$ for ${U>0}$ and $\tilde{\alpha}=\alpha^{\prime}$ for $U<0$. 
Note that, in Eq. (\ref{weightsigma}) the notion $\sigma$ is introduced in $H^{\sigma}_{G,D,S}$. Without Rashba spin-orbit coupling, the variants of the KM models still preserve $s_z$, such that $H^{\sigma}_{G,D,S}$ are decoupled for different spin flavors.
For the $U>0$ case, it is easy to show that at half-filling, PHS in the variants of the KM model renders $w_{\uparrow}w_{\downarrow}=|w_{\uparrow}|^2>0$, such that the Monte Carlo simulations always have positive-definitive sampling.\cite{zheng2011,meng2013} For the $U<0$ case, the time-reversal symmetry guarantees $w_{\uparrow}=w^*_{\downarrow}$, so is  the positiveness of $w_{\uparrow}w_{\downarrow}$. Therefore, we can have sign-free simulations and numerically exact solutions for the repulsive GKM, DKM-Hubbard model and attractive SKM-Hubbard model.

It has been pointed out by a QMC study\cite{Hung2013} that it is not easy to get access to the interacting topological phase transition by identifying a gap closing, due to strong finite-size effect in single-particle gaps. The straightforward approach to determine locations of the topological phase transition boundaries is to evaluate the $Z_2$ topological index and spin Chern number, in terms of zero-frequency Green's functions.\cite{Hung2013,WangPRB,Wang_PRX,Wang2013,meng2013,hung2014} With the inversion symmetry, the interacting $Z_2$ invariant $\Delta$ is evaluated as
\begin{equation}
(-1)^{\Delta}=\prod_{\textrm{R-zeros}} \xi^{1/2}({\bf \Gamma}_i),
\end{equation}
where  $\xi({\bf \Gamma}_i)$ denotes the parity eigenvalue of the "R-zeros" eigenstates of interacting zero-frequency Green's function at TRIM points ${\bf M}_i$ [here in the KM model, the TRIM  are ${\bf \Gamma}=(0,0)$ and ${\bf M}_{1,2,3}$ as shown in Fig. \ref{Fig:BZ} .]
More explicitly, 
\begin{equation}
G(i\omega=0,{\bf \Gamma}_i )|\mu\rangle= \mu |\mu \rangle,
\end{equation}
with $\mu>0$ and $P|\mu\rangle=\xi({\bf \Gamma}_i)|\mu \rangle$, where $P$ is the parity operator. The zero-frequency Green's functions are obtained through time-displaced Green's functions in the QMC simulations and with Fourier transformation. More detailed procedures has been indicated in the review article\cite{meng2013}.

\begin{figure}[t!]
\centering
   \subfigure{\includegraphics[width=2.4 in]{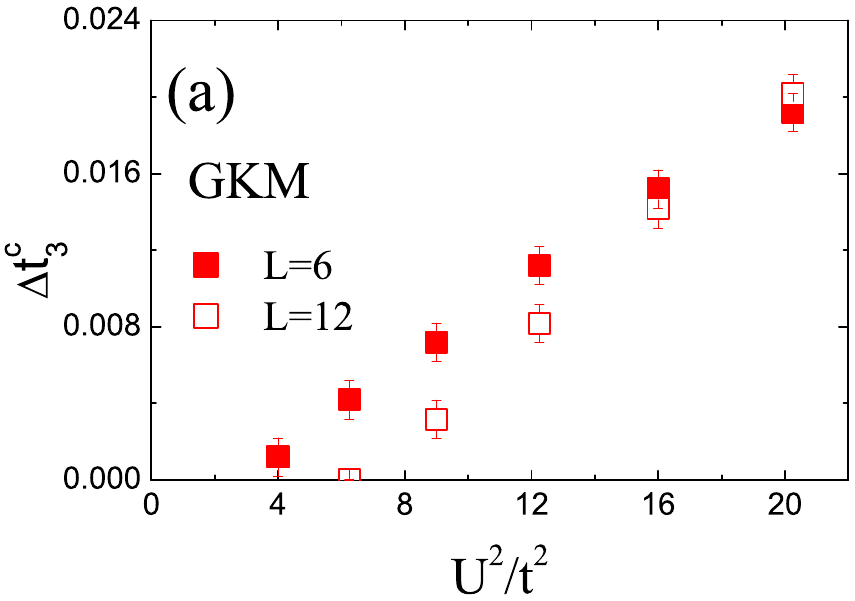}
      \label{Fig:GKM_QMC}}
   \subfigure{\includegraphics[width=2.4 in]{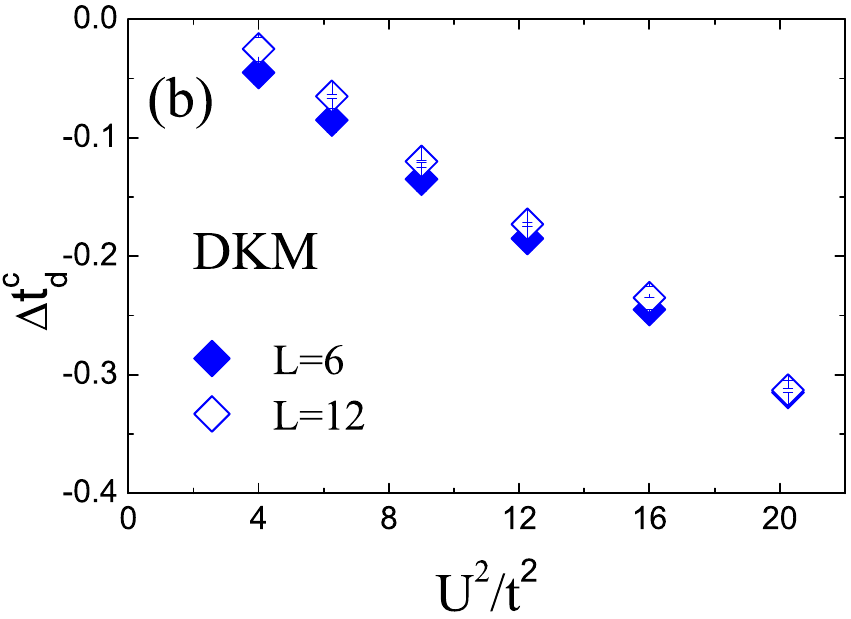} 
     \label{Fig:DKM_QMC}}
\caption{QMC data for QSH boundary shift (a) $\Delta t^c_3=t^c_3(U)-\frac{1}{3}t$ in the GKM-Hubbard model, and (b) $\Delta t^c_d=t^c_d(U)-2t$ in the DKM-Hubbard model. In (a), the shift is positive, which means QSH is more stable against $t_3$ hoppings. More interestingly, the shift amount is linearly proportional to $(U/t)^2$, consistent with our mean-field picture. In (b), the shift is negative and linearly proportional to $(U/t)^2$. The short-range interaction makes the QSH phase more destabilized by the dimerization $t_d$. Statistical errors are denoted by the error bars.}
\label{Fig:QMC}
\end{figure}

The numerical results are shown in Fig.~\ref{Fig:QMC}. The closed (open) red squares and blue diamonds represent the boundary shifts $\Delta t^c_3=t^c_3(U)-\frac{1}{3}t$ and $\Delta t^c_d=t^c_d(U)-2t$ obtained in the GKM-Hubbard and DKM-Hubbard modes on $6\times 6$ ($12\times 12$) clusters, respectively. Here we consider the discretized time step $\Delta \tau t=0.05$. In both models, we sweep the critical points $t^c_3$ and $t^c_d$ at a variety of $U$, (the strength of $U$ is below the critical value to magnetic instability). The topological phase transitions appear when the interacting $Z_2$ invariant has changes  $\Delta=-1 \leftrightarrow1$. In both models, the QMC results show that appropriate sign and amounts of the boundary shift are linearly proportional to $(U/t)^2$ to high accuracy, as the PMF theory predicted. This means that the PMF can properly capture the correlation effect at the intermediate interaction realm. Note that the linear relations to $(U/t)^2$ are universal and size-independent in the QMC results. Compared with Figs. \ref{Fig:GKMDKM}, one can see that the PMF theory has captured and interpreted the behavior.

\begin{figure}
\centering
 \includegraphics[width=2.5 in]{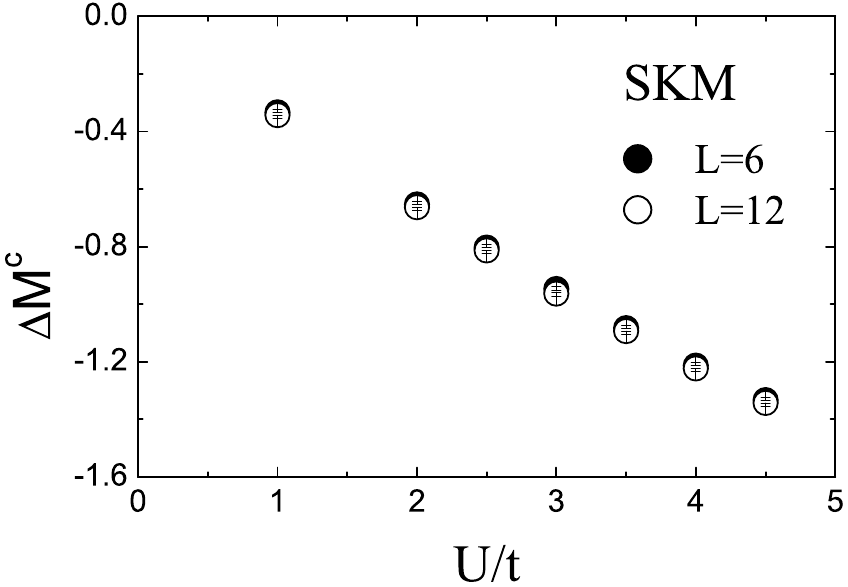}
\caption{QMC data for QSH boundary shift in the attractive Kane-Mele-Hubbard model in the presence of staggered potential $M$. The shift amount of the critical $M$, $\Delta M^c=M^c(U)-\sqrt{3}\lambda_{SO}$ is negative and linearly proportional to $U/t$. Note that we chose $\lambda_{so}=0.2t$ and $U<0$. PHS is explicitly broken due to the staggered potentials. QMC is only fermion-sign free in the attractive $U<0$ case. }
\label{Fig:QMC_staggerKM}
\end{figure}

Next we move to the SKM-Hubbard model. Due to the absence of PHS, the QMC simulation on the repulsive SKM-Hubbard model has minus sign problems. Instead, we  turn to study  the {\it attractive} staggered Kane-Mele-Hubbard model ($U=-|U|<0$ and $\lambda_{so}=0.2t$) on $6\times 6$ and $12\times 12$ sites, and depict the topological boundary shift under correlation in Fig.~\ref{Fig:QMC_staggerKM}. The QMC results also explicitly support the PMF's prediction, that the topological phase boundary is linearly proportional to the first-order $U/t$, and the negative slope of the $M^c-U$ curve, indicating that the attractive interaction destabilizes the topological phase. For other KM-type models, we believe our approach in this work, $U/t$ expansion $+$ mean-field treatment $+$ low-energy gap equation $\Delta({\bf k})$, can essentially capture the interactions effects on the $Z_2$ topological phase transitions, and possibly in more general models as well.

\section{Discussion}\label{Sec:Discussion}
Readers may be confused about why the boundary shifts due to the presence of the onsite interaction can not be studied by perturbative methods such as weak-coupling RG analysis \cite{Shankar_RGRMP} at the critical phase in which the gaps close to form Dirac points. We note that such RG analysis can not predict any phase boundary shift. For simplicity and illustration, we focus on the critical phase (semi-metal phase) at the phase transition in SKM below in Sec.~\ref{Subsec:RG_critical}, where the number of relevant degrees of freedom is smaller compared with those in GKM and DKM due to the presence of more number of Dirac nodes at the critical phase. The straightforward thought is that even though the local four-fermion interactions are irrelevant in the critical phase, before they flow to negligible values under RG it is possible that they still generate a small bilinear mass term, which can possibly shift the phase boundary. However, below we will explicitly perform tree-level RG analysis and show that the tree-level RG corrections completely cancel each other, which gives no generation of a bilinear mass term. Hence we conclude that the boundary shift can only be captured by the physics of the lattice model and can not be captured by the coarse-grained continuum theory around ${\bf K}$ and ${\bf K'}$. 

We note that the reason that we consider the tree-level RG analysis is to see if the four-fermion terms at low-energy limit can generate an infinitesimal bilinear term that can shift infinitesimally the critical point. Of course, we can consider the one-loop RG corrections and see the effects but some concerns should be kept in mind. First, the one-loop RG equations schematically will contribute to the RG equation as 
\begin{eqnarray}
\frac{dg}{d\ell} \simeq -g + \alpha g^2,
\end{eqnarray}
where $\alpha$ is some finite constant and $g$ is the coupling strength of the four-fermion term associated with the local interaction. We can clearly see there exists a quantum critical point dictating the phase transition between the semimetal phase and some symmetry broken phase. However, this analysis does not provide any information about the shift of the boundary between the topological phase and the topologically trivial phase. Second, in order for the occurrence of the phase transition, the bare value of the short-range coupling $g$ is actually of order $O(1)$ which is beyond the weak-coupling regime. This analysis implies there is a critical point but the explicit result may be controversial and needs checking by other approaches. 

In Sec.~\ref{Subsec:UV}, we will consider a more extended interaction including both Hubbard $U$ and nearest neighbor interaction $V$ in the SKM within the self-consistent PMF. The inclusion of the nearest-neighbor $V$ complicates the analysis. We find that whether or not the topological phase is stabilized due to the short-ranged interactions depends on the details of the competition between the onsite $U$ and the nearest-neighbor $V$. From the low energy analysis focusing on ${\bf K}$ point in the B.Z., we conclude that within the mean-field picture if $U$ is dominant over $V$ ($U > 6V$ ) the qualitative result obtained from the case with only onsite $U$, repulsive (attractive) interactions stabilize (destabilize) the topological phase, is still correct in the $U$-$V$ model. However, from the studies of the Kane-Mele-$U$-$V$ model, it may suggest a long-ranged repulsive interaction such as Coulomb interaction may completely destabilize the QSH phase, which is, however, contrary to the recent QMC studies\cite{Hohenadler_KMC}. The contradiction is due to the underestimation of the effects of the ``long-ranged tail`` of the Coulomb interaction. 

\subsection{RG analysis of the critical phase in the SKM model with weak interaction $U$}\label{Subsec:RG_critical}
In this model, since $S^z$ is still conserved, the spin-up and spin-down Hamiltonian can be treated separately. For each spin species, we can diagonalize the Hamiltonian matrix for spectra. There are $2$ bands for each spin species. The bands can be characterized by the eigenvector-eigenenergy pairs $\{ \vec{v}^\alpha_{b}({\bf k}),\epsilon^\alpha_{b} ({\bf k})\}$, where $b=1,2$ are band indices. The Hamiltonian can be diagonalized by rewriting the original fermion fields in terms of the complex fermion fields $d^\alpha_b({\bf k})$ in the diagonal basis,
\begin{eqnarray}
c_{\alpha}({\bf r},a) = \sqrt{\frac{1}{N_{uc}}}\sum_{b=1,2}\sum_{{\bf k}\in{\bf B.~Z.}}v^\alpha_{b}({\bf k},a) d^\alpha_b ({\bf k}) e^{i {\bf k}\cdot {\bf r}},~~
\end{eqnarray}
where $N_{uc}$ is the number of unit cells and the complex fermion field $f$ satisfies the usual anticommutation relation $ \{ d^{\alpha \dagger}_{b}({\bf k}), d^{\alpha'}_{b' }({\bf k'}) \} = \delta_{\alpha \alpha'}\delta_{b b'} \delta_{{\bf k} {\bf k'}}$. In terms of the new complex fermion fields, the Hamiltonian becomes
\begin{eqnarray}
H_{KMs} = \sum_{b=1,2} \sum_{\alpha=\uparrow,\downarrow} \sum_{{\bf k}\in{\bf B.~Z.}}\epsilon^\alpha_{b}({\bf k}) d^{\alpha\dagger}_b({\bf k}) d^\alpha_b ({\bf k}).
\end{eqnarray}
At the critical phase $(\lambda^c_m = 3\sqrt{3} \lambda_{so})$, the gaps close at momentums ${\bf K}$ and ${\bf K'} = - {\bf K}$. Around these points, only the spin-down fermions are gapless at ${\bf K}$ and spin-up fermions are gapless at ${\bf K'}$. As far as the long-wavelength (low-energy) description is concerned, we can focus on the ${\bf K}$ and ${\bf K'}$ points and perform expansion around theses points by introducing a small momentum shift $\delta{\bf k}$.

For the low-energy description at momentum ${\bf K}$, we find that only the spin-down fermions are gapless and the expansion around ${\bf K}$ by introducing ${\bf k} = {\bf K} + \delta{\bf k}$, with $| \delta {\bf k}| <\Lambda,~\Lambda \ll |{\bf K}|$, gives
\begin{eqnarray}
H_{\bf K} \simeq \sum_{|\delta {\bf k}|<\Lambda} v_F |\delta{\bf k}| \bigg{[} && \psi^\dagger_{1R \downarrow}(\delta{\bf k}) \psi_{1R\downarrow}(\delta {\bf k})  - \psi^\dagger_{2R\downarrow}(\delta {\bf k}) \psi_{2R\downarrow}(\delta {\bf k})\bigg{],}\label{Eq:H_K}
\end{eqnarray}
where we introduced $d^\downarrow_b({\bf K} + \delta {\bf k}) \equiv \psi_{bR\downarrow}(\delta {\bf k})$  with $R$ labeling the valley at ${\bf K}$ and $v_F \equiv \sqrt{3}t/2$ is the Fermi velocity of each band at ${\bf K}$. It is more convenient to transform the continuum fields defined above to real space, defining
\begin{eqnarray}
\psi_{bR\downarrow}({\bf r}) =\sqrt{\frac{1}{N_{uc}}} \sum_{|\delta{\bf k}|< \Lambda} e^{i \delta{\bf k}\cdot {\bf r}}\psi_{bR\downarrow}(\delta{\bf k}).
\end{eqnarray}
Therefore, in the low-energy description, we can re-express the spin-down fermion field as
\begin{eqnarray}
c_{\downarrow}({\bf r},a) \simeq \sum_{b=1,2} v_{b R\downarrow}(a) \psi_{b \downarrow}({\bf r})e^{i {\bf K}\cdot {\bf r}},
\end{eqnarray}
where we defined $v^{\downarrow}_b ({\bf K}+ \delta{\bf k},a) \equiv v_{bR\downarrow}(a)$.

Similarly, we can also obtain the low-energy description at ${\bf K'}$. At ${\bf K'}$, only the spin-up fermions are gapless and expansion around ${\bf K'}$ with small momentum shift $\delta {\bf k}$ gives
\begin{eqnarray}
H_{\bf K'} \simeq \sum_{\delta {\bf k}<\Lambda} v_F |\delta {\bf k}|\bigg{[} &&\psi^\dagger_{1 L\uparrow}(\delta{\bf k}) \psi_{1L\uparrow}(\delta {\bf k})  - \psi^\dagger_{2L\uparrow}(\delta {\bf k}) \psi_{2L\uparrow}(\delta {\bf k})\bigg{]}, \label{Eq:H_K'}
\end{eqnarray}
where similarly we defined $d^\uparrow_b({\bf K'} + \delta{\bf k}) \equiv \psi_{bL\uparrow}(\delta{\bf k})$. We can define a similar transformation to the real space as above, and the spin-down fermion field can be effectively expressed as
\begin{eqnarray}
c_{\uparrow}({\bf r},a) \simeq \sum_{b=1,2} v_{b L\uparrow}(a)\psi_{bL\uparrow}({\bf r})e^{i{\bf K'}\cdot {\bf r}},
\end{eqnarray}
with $v^\uparrow_b({\bf K'} + \delta {\bf k},a) \equiv v_{b L\uparrow}(a)$, $L$ labeling the valley at ${\bf K'}$, and remember ${\bf K'} = - {\bf K}$. The action for the low-energy description is
\begin{eqnarray}
S_{0,P} = \int \frac{d^2{\bf q} d\omega}{(2\pi)^3} \left[\psi^\dagger_{b P\alpha_P} (q) (-i \omega) \psi_{b P\alpha_P} (q) + H_P\right],~~~~~
\end{eqnarray}
with $P = R/L = {\bf K}/{\bf K'}$ and $\alpha_{R/L} = \downarrow/\uparrow$ and we use $2+1$ dimensional vector $q$ representing frequency and momentum $(\omega,~{\bf q})$. We can also define the Green's functions as
\begin{eqnarray}
 \la \psi^\dagger_{bL\uparrow}(q) \psi_{bL\uparrow}( q') \ra = \la \psi^\dagger_{bR\downarrow}( q) \psi_{bR\downarrow}(q') \ra = \frac{i \omega -(-1)^b v_F |{\bf q}|}{(i\omega)^2 - (v_F |{\bf q}|)^2} \delta^{(3)}_{q q'},
\end{eqnarray}
and we introduce the abbreviation $\delta^{(3)}_{q q'} = (2\pi)^3 \delta(\omega-\omega') \delta^{(2)}({\bf q} - {\bf q'})$.

In order to write down the general expression of the four-fermion interactions, we need first to obtain the symmetry transformation of the fields defined above. There are $S^z$-conservation, $U(1)$-charge, TRS, and $C_3$ in this system. Except TRS, the other else symmetry transformations are quite transparent. Let's focus on the symmetry transformation under TRS ($\mathcal{T}$), and we find
\begin{eqnarray}
&& v^{\uparrow *}_{bL}(a)  \mathcal{T} \psi^\uparrow_{bL} \mathcal{T}^{-1} = -v^\downarrow_{bR}(a)\psi^\downarrow_{bR},\label{Eq:TRS_1}\\
&& v^{\downarrow *}_{bR}(a) \mathcal{T} \psi^{\downarrow}_{bR} \mathcal{T}^{-1} =v^\uparrow_{bL}(a)\psi^\uparrow_{bL}.\label{Eq:TRS_2}
\end{eqnarray}
With TRS, the eigenvector-eigenvalue pairs have the property, $\vec{v}^\uparrow_b ({\bf k})=[\vec{v}^\downarrow_b (-{\bf k})]^*$  and $\epsilon^{\uparrow}_b ({\bf k}) = \epsilon^{\downarrow}_b (-{\bf k})$, which gives $v^{\uparrow *}_{b L}(a) = v^{\downarrow}_{bR}(a)$. We can use the properties above to simplify the TRS transformation in (\ref{Eq:TRS_1}) and (\ref{Eq:TRS_2}), but as far as the RG analysis presented below is concerned, we don't need to do that.

The general expressions of the local four-fermion interactions in terms of the continuum fields defined above are shown below. For simplicity in the expression, we define below $f_{b P \alpha}(a) \equiv v^\alpha_{b P} (a) \psi^\alpha_{b P}( {\bf r})$, and the local four-fermion action can be written as (repeated $a$ means summation over the eigenvector elements)
\begin{eqnarray}
\nonumber S_{int} =~&& \omega^{a}_{11} f^\dagger_{1L\uparrow}(a) f_{1L\uparrow}(a) f^\dagger_{1R\downarrow}(a)f_{1R\downarrow}(a) + \omega^a_{22} f^\dagger_{2L\uparrow}(a) f_{2L\uparrow}(a) f^\dagger_{2R\downarrow}(a) f_{2R\downarrow}(a) +\\
\nonumber && + \omega^a_{12}\left[f^\dagger_{1L\uparrow}(a)f_{1L\uparrow}(a)f^\dagger_{2R\downarrow}(a)f_{2R\downarrow}(a) + f^\dagger_{1R\downarrow}(a)f_{1R\downarrow}(a)f^\dagger_{2L\uparrow}(a) f_{2L\uparrow}(a)\right]+\\
\nonumber && +\lambda^a_{12}\left[ \left( f^\dagger_{1L\uparrow}(a)f_{1L\uparrow}(a)f^\dagger_{1R\downarrow}(a)f_{2R\downarrow}(a) +f^\dagger_{1R\downarrow}(a)f_{1R\downarrow}(a)f^\dagger_{1L\uparrow}(a)f_{2L\uparrow}(a)\right)+\Hc \right] + \\
\nonumber && + \lambda^a_{21}\left[\left( f^\dagger_{2L\uparrow}(a) f_{2L\uparrow}(a)f^\dagger_{1R\downarrow}(a)f_{2R\downarrow}(a)+f^\dagger_{2R\downarrow}f_{2R\downarrow}(a)f^\dagger_{1L\uparrow}(a)f_{2L\uparrow}(a)\right)+\Hc \right] +\\
\nonumber && + u^a_{12}\left[ f^\dagger_{1L\uparrow}(a) f_{2L\uparrow}(a)f^\dagger_{1R\downarrow}(a)f_{2R\downarrow}(a) + \Hc \right]  +\\
&&+ u^a_{21} \left[ f^\dagger_{1L\uparrow}(a)f_{2L\uparrow}(a) f^\dagger_{2R\downarrow}(a)f_{1R\downarrow}(a) + \Hc \right],
\end{eqnarray}
and we remark that in the presence of the onsite interaction $U$, all the bare couplings above are simply equal to $U$.

The RG analysis in a nutshell is first to integrate out the fast-momentum modes defined within a momentum shell between $[\Lambda/b, \Lambda]$, with $b \equiv e^{d\ell}\simeq 1 + d\ell$ slightly bigger than one. Then we rescale the (fermion or boson) fields and the momentum to recast the action back to the original form. After the elimination of the fast-momentum modes and rescaling process, we can examine how the couplings of the interactions change. The mathematical form at the tree-level is
\begin{eqnarray}
S_{eff,<} = \la S_{int} \ra_>,
\end{eqnarray}
where the subscript $>$ means momentum shell integral of the fast-momentum modes.

At the tree-level, we find the corrections are
\begin{eqnarray}
\nonumber \la S_{int} \ra_>=  &&-\frac{\Lambda^2}{4\pi}d\ell  \bigg{[} \omega^a_{11} \big{|}v_{1R\downarrow}(a) \big{|}^2 - \omega^a_{12} \big{|} v_{2R\downarrow}(a) \big{|}^2 \bigg{]}f^\dagger_{1L\uparrow}(a)  f_{1 L \uparrow}(a)  +\\
\nonumber && + \frac{\Lambda^2}{4\pi}d\ell \bigg{[} \omega^a_{22} \big{|} v_{2R\downarrow}(a) \big{|}^2 - \omega^a_{12} \big{|} v_{1R\downarrow}(a) \big{|}^2 \bigg{]} f^\dagger_{2L \uparrow}(a) f_{2L\uparrow}(a) -\\
\nonumber && - \frac{\Lambda^2}{4\pi}d\ell \bigg{[} \omega^a_{11} \big{|} v_{1L\uparrow}(a) \big{|}^2 - \omega^a_{12} \big{|} v_{2L \uparrow}(a) \big{|}^2 \bigg{]} f^\dagger_{1R \downarrow}(a) f_{1R\downarrow}(a) + \\
\nonumber && + \frac{\Lambda^2}{4\pi}d\ell \bigg{[} \omega^a_{22} \big{|} v_{2L \uparrow}(a) \big{|}^2 - \omega^a_{12} \big{|} v_{1L\uparrow}(a) \big{|}^2 \bigg{]} f^\dagger_{2R\downarrow}(a) f_{2R\downarrow}(a)-\\
\nonumber && -\frac{\Lambda^2}{4\pi}d\ell  \bigg{[} \lambda^a_{12}\big{|} v_{1L\uparrow}(a) \big{|}^2  -  \lambda^a_{21} \big{|}v_{2L\uparrow}(a) \big{|}^2\bigg{]}\left( f^\dagger_{1R\downarrow}(a) f_{2R\downarrow}(a)+\Hc \right)-\\
&& -\frac{\Lambda^2}{4\pi} d\ell \bigg{[}  \lambda^a_{12}\big{|}v_{1R\downarrow}(a) \big{|}^2 -  \lambda^a_{21}\big{|} v_{2R\downarrow}(a) \big{|}^2\bigg{]}\left(f^\dagger_{1L\uparrow}(a) f_{2L\uparrow}(a)+ \Hc \right),~~ \label{Eq:tree-level_RG}
\end{eqnarray}
and all the four-fermion couplings are irrelevant,
\begin{eqnarray}
\frac{dg}{d\ell} = - g,
\end{eqnarray}
with $g= \omega^a$-s, $\lambda^a$-s, $u^a$-s introduced above and $\ell$ is the logarithm of the length scale in RG analysis..

The bare couplings of $\omega^a_{11}(\ell =0) = \omega^a_{22}(0) = \omega^a_{12} (0)= U$, and we numerically check that $ \big{|} v_{1R\downarrow}(a) \big{|}^2 = \big{|} v_{2R\downarrow}(a) \big{|}^2$, $\big{|} v_{1L\uparrow}(a)\big{|}^2 = \big{|} v_{2L\uparrow}(a)\big{|}^2$. At the tree-level RG analysis, all the couplings decays at the same rate under RG flow. Before all the four-fermion couplings flow to negligible values, at some small $\ell_c$, we have $\omega^a_{11} (\ell_c) = \omega^a_{22}(\ell_c) =\omega^a_{12}(\ell_c)=\lambda^a_{12}(\ell_c) = \lambda^a_{21}(\ell_c)$, and hence the bilinear corrections generated by these irrelevant four-fermion interactions completely cancel each other, which {\it leaves no corrections at the tree-level RG analysis.} Therefore, we conclude such long-wavelength analysis can not capture the shift of the boundary between the topologically trivial and nontrivial phases. The boundary shift can only be captured, at least in this model, by the {\it lattice} Hamiltonian which is not coarse-grained.

\subsection{SKM-U-V model}\label{Subsec:UV}
We will consider a more extended interaction which includes both the Hubbard $U$ and the nearest-neighbor $V$. The presence of the nearest-neighbor $V$ within mean-field treatment contribute both the diagonal terms and the off-diagonal terms to the original Hamiltonian. The diagonal terms obviously renormalize the mass terms and the off-diagonal terms renormalize the nearest-neighbor hopping amplitude $t$, resulting in renormalizing the velocity of the Dirac fermions in the critical phase. Within the mean-field picture, besides the expectation values of the densities defined in the onsite Hubbard case, we also need to introduce
\begin{eqnarray}
&& \bigg{\la} c^\dagger_{\sigma}({\bf r}, B) c_\sigma({\bf r},A) \bigg{\ra} \equiv \left( \chi_\sigma \right)^*,\\
&& \bigg{\la} c^\dagger_{\sigma}({\bf r} + {\bf e}_a, B) c_\sigma({\bf r},A) \bigg{\ra} \equiv \left( \chi_{\sigma}({\bf e}_a)\right)^*,
\end{eqnarray}
where ${\bf e}_{a = 1,2}$ defined in Fig.~\ref{Fig:honeycomb}. The nearest-neighbor $V$ contributes additional terms to the full Hamiltonian. In the matrix form, the additional terms can be expressed as
\begin{equation}
h_v({\bf k})=V \begin{pmatrix}
3\la n_B \ra & -\frak{f}_{\uparrow}({\bf k}) & 0 & 0\\
-(\frak{f}_{\uparrow}({\bf k}))^* & 3 \la n_A \ra & 0 & 0\\
0 & 0 & 3 \la n_B \ra & - \frak{f}_{\downarrow}({\bf k})\\
0 & 0 & - ( \frak{f}_{ \downarrow}({\bf k}) )^* & 3 \la n_A \ra
\end{pmatrix},
\end{equation}
where $\frak{f}_{\sigma}({\bf k}) \equiv (\chi_{\sigma} )^* + e^{i {\bf k}\cdot {\bf e}_1} (\chi_\sigma ({\bf e}_1))^* + e^{i {\bf k}\cdot {\bf e}_2} (\chi_\sigma ({\bf e}_2))^*$. By $C_3$ symmetry, we can simplify the result by identifying $\chi_\sigma = \chi_\sigma({\bf e}_1) = \chi_\sigma({\bf e}_2)$. We can see that $f_{\sigma}({\bf k})$ is proportional to $f({\bf k})$ defined in Eq.~(\ref{Eq:SKM_H}) and therefore vanish at momentums ${\bf K}$ and ${\bf K'}$. 
\begin{figure}[t]
\centering
\subfigure[]
{\label{Fig:UV_repulsive} \includegraphics[width=2.4in]{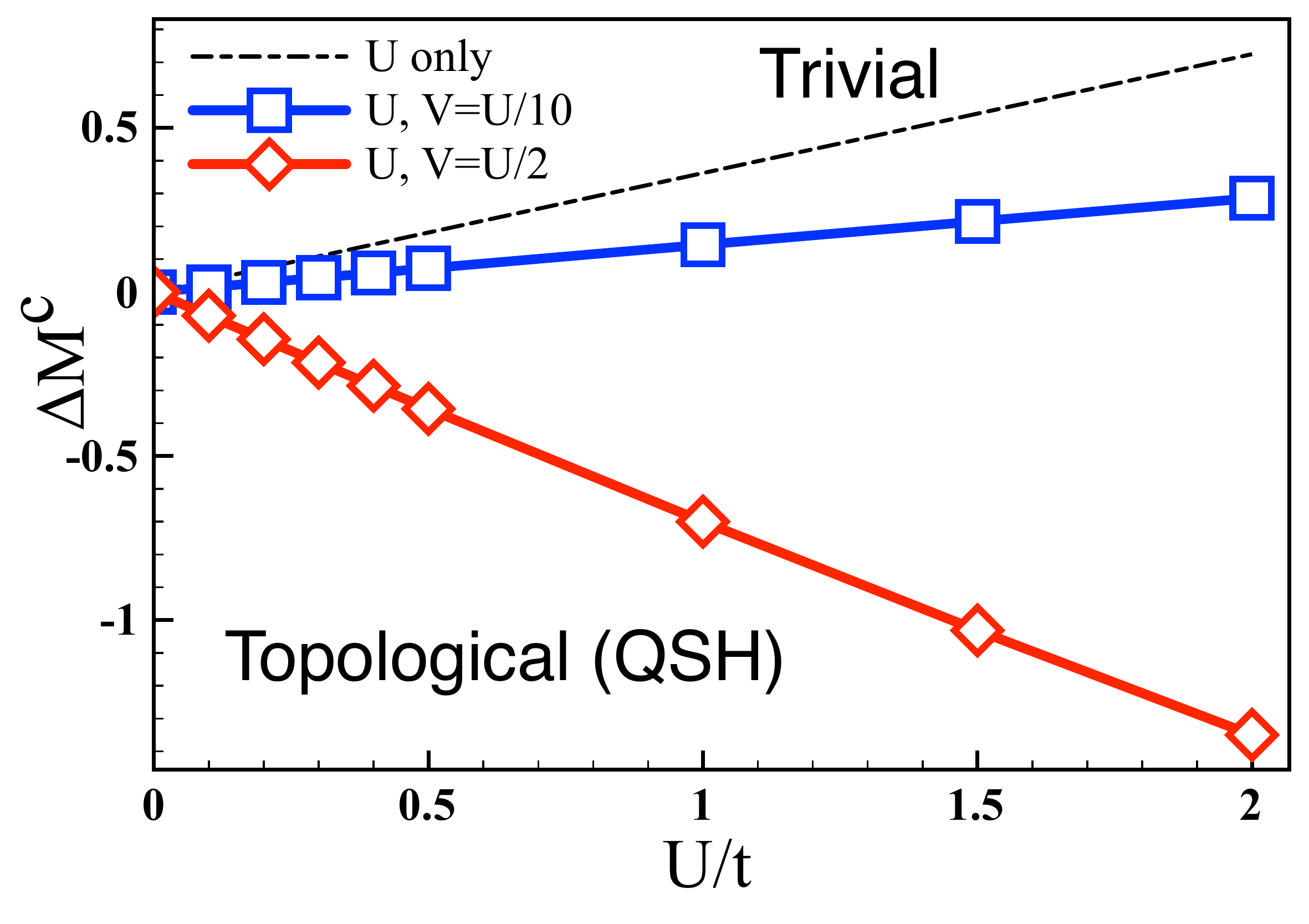}}
\subfigure[]
{\label{Fig:UV_attractive} \includegraphics[width=2.4 in]{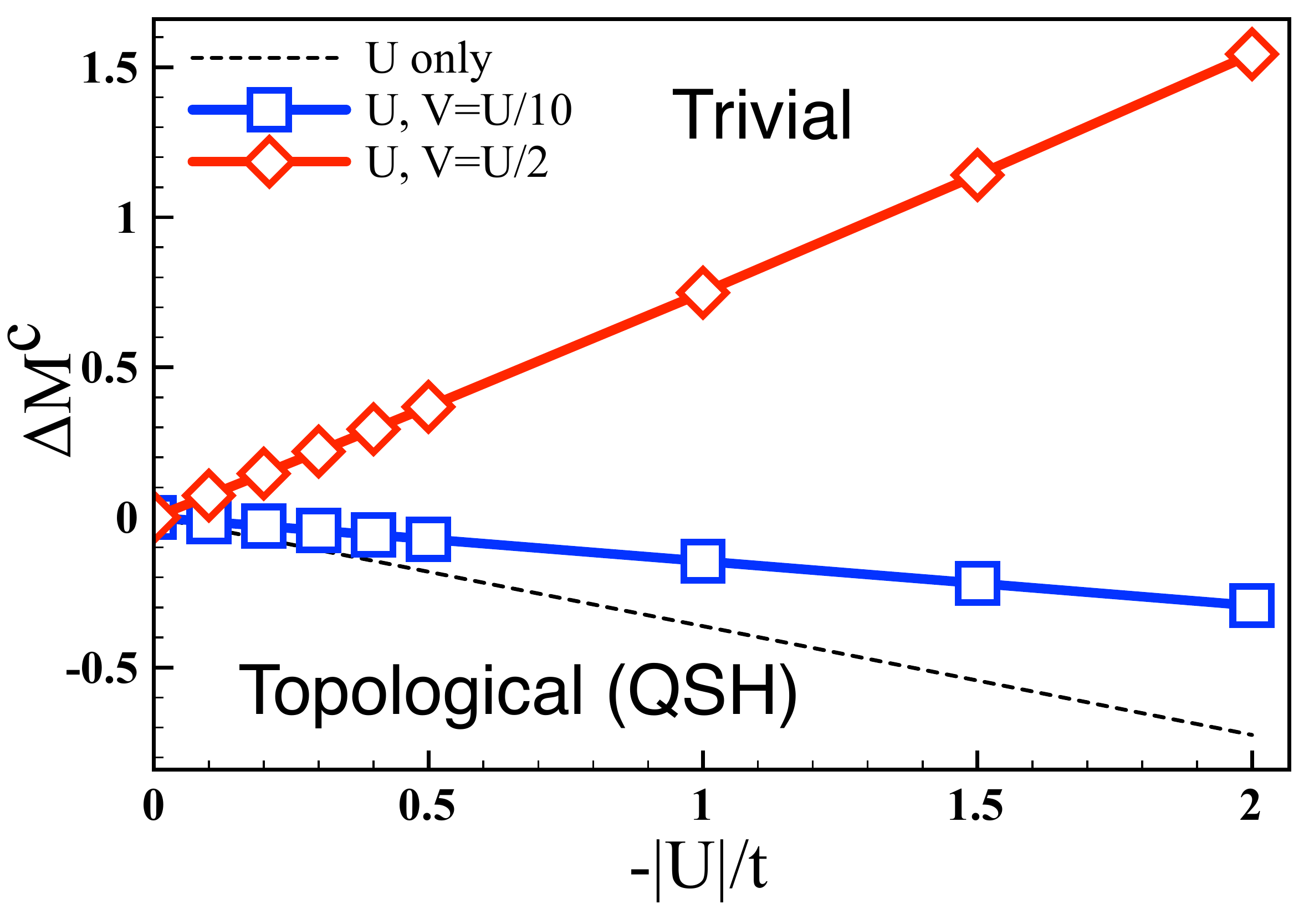}}
\caption{Illustration of the phase boundary shift in the present of both U and V repulsion. We take $t=1$, $\lambda_{so} =0.4$, and choose $V=U/2$ and $V=U/10$. The blue open dot squares represent the boundary in the case of $V=U/10$ and the open red diamonds line represents the boundary in the case of $V=U/10$. The black line represents the boundary in the present of onsite Hubbard $U$. (a) The repulsive interaction case, $U, V >0$. We can see that in this case. If the nearest-neighbor $V$ is much smaller than $U$, $V=U/10$, the interactions still tend to stabilize the topological phase. But due to the competition between $U$ and $V$ the topological widow is less widened by the interactions. If $V$ is more comparable to $U$,$V=U/2$ in this case, the effects of $V$ will be dominant over $U$ and will tend to destabilize the topological phase. According to the gap function defined around momentum ${\bf K}$ point, Eq.~(\ref{Eq:UV_gap}), the transition point is at $U= 6V$. (b) The attractive interaction case, $U, V <0$. The results in this case are qualitatively opposite to those in the repulsive case. When $V$ is more comparable to the $U$, the interactions tend to stabilize the topological phase.}
\label{Fig:UV_data}
\end{figure}
Focusing on momentum ${\bf K}$, we can see that the two of the four bands with the eigenvalues $E_{1/2}=\pm M \mp 2\lambda_{so}g(K) + \frac{U}{2} \la n_{A/B} \ra + 3V\la n_{B/A} \ra$ can be inverted due to the tuning of the ratio of $M$ and $\lambda_{so}$. Therefore, we define the gap function as
\begin{eqnarray}
\nonumber && \Delta({\bf K}) = E_1 - E_2 \\
 && =  2 M - 4 \lambda_{so} g(K) + \left( \frac{U}{2}-3V\right) \big{[} \la n_A \ra - \la n_B \ra \big{]}.\label{Eq:UV_gap}
\end{eqnarray}
Due to the presence of the staggered potentials, the density at B is larger than that at A, $\la n_B \ra > \la n_A \ra$, and the sign of the correction of the last term in the gap function depends on the competition between $U$ and $V$. For repulsive $U, V >0$, if $U>6V$, the last term is negative and the repulsive interactions stabilizes the topological phase. However, if $U<6V$, the last term is positive and then the interactions destabilizes the topological phase. On the other hand, the attractive $U, V <0$ would give the opposite results to the repulsive case.

For illustration, we numerically check the cases with $t =1$, $\lambda_{so} =0.4$, and choose $V=U/10$ and $V=U/2$ on a honeycomb lattice consisting of $200\times 200$ unit cells. The results are shown in Fig.~\ref{Fig:UV_data}. The blue open dot squares represent the topological phase boundary shift in the case of $V=U/10$ and the open red diamonds line represents the boundary shift in the case of $V=U/10$. The black dashed line represents the boundary shift in the present of only onsite Hubbard $U$ shown in Fig.~\ref{Fig:ShiftSKM}. Qualitatively, the results between the repulsive and the attractive case are opposite. In the repulsive interaction case, the more short-ranged the repulsive interactions are, the more stable the topological is. On the other hand, in the attractive interaction case, the more extended the attractive interactions are, the {\it more stable} the topological phase is. In the end, we remark that the extended repulsion/attraction also enhance the fluctuations of charge/spin which lead to the magnetic phase transition. The extended part of the interaction tends to ``decrease`` the onsite repulsive interaction. The results will lead to the modification of the familiar two spin exchanges in the large $U$ limit, $J_{\bf r r'} = 4t^2_{\bf r r'}/U \rightarrow 4t^2_{\bf r r'} / ( V_0 - V_{{\bf r}- {\bf r'}})$, where we define $V_0$ corresponds to onsite Hubbard reuplsion and $V_{{\bf r} - {\bf r'}}$ corresponds to the extend part. Similarly, all the higher order multispin excahnge amplitudes are modified and may in fact be relatively more important in systems with extended interactions.\cite{Schuler2013} Therefore, the magnetic phase transition may occur in a larger value of the onsite interaction strength, which will make the topological phase boundary shifts more obvious.

\section{Conclusion and Outlooks}\label{Sec:Conclusion}
The main object of this short review is to introduce an analytical method combining perturbation and self-consistent mean-field treatments which can be used to study how a local interaction renormalizes the critical value of the parameter that drives a $Z_2$ topological transition. For demonstration of the validity of our method, we illustrate our method on three different variants of the Kane-Mele-Hubbard models, which were previously well-studied numerically, and we find the signs of the shifts and scalings of topological phase boundary are in excellent agreement with previous quantum Monte-Carlo simulations. We conclude that the shift amounts of the QSH boundary in the generalized Kane-Mele and dimerized Kane-Mele models are proportional to $(U/t)^2$, while that in the stagger-potential Kane-Mele model is linearly proportional to $U/t$.

The analytical approach introduced in this review, we believe, is the simplest way to explain the shift of the critical parameter value due to the interaction effects in the parameter-driven topological phase transition, and has general applicability. In two dimensions, the consistency between the analytical results and the exact quantum Monte-Carlo results in the Kane-Mele-Hubbard model with particle-hole symmetry gives us somewhat firm ground. We can certainly to apply the analytical method to analyze other lattice models with \cite{hohenadler2013} or without particle-hole symmetry \cite{Miguel2013} in which exact quantum Monte-Carlo approach suffers the sign problem.

This approach especially can benefit most the analysis in certain three-dimensional topological phase transitions where the exact QMC simulations suffer from the sign problem and the system size limitation. As far as the model studies are concerned, the simplest three-dimensional model, to our knowledge, which can host $Z_2$ topological phase transitions is the tight-binding-type model with spin-orbit couplings on the diamond lattice proposed by Fu, Kane, and Mele (FKM).\cite{FKM_model} The noninteracting FKM model with an anisotropic hopping bond can host a topological transtion from topological insulator to trivial insulator through a gap-closing process. The analytical approach developed in this review can be directly applied to analyze the interacting correlation effects on the shift of the critical parameter value in the FKM model. 

Furthermore, the FKM model in the presence of staggered potentials are also studied recently. \cite{ojanen2013} In the presence of the staggered potentials, the inversion symmetry is explicitly broken and the Weyl semimetal phase is realized. In this case, there exists a critical strength of the staggered potential, after which the Weyl nodes and the anti-Weyl nodes annihilate each other and there is a transition from a Weyl-semimetal to a trivial insulator. We speculate that the analytical method introduced in this review article is applicable to examine how the critical strength of the staggered potential gets renormalized due to the presence of the short-range interaction. In sum, we believe our method is applicable to rather general systems with local interactions.  Our approach helps to enrich the overall picture of how interactions influence topological phases, with the potential to help materials scientists ``engineer" new platforms (beyond the familiar suite of topological insulators) for technologies based on TI (including those groups working on a cold-atom approach to TI).

\section*{Acknowledgments}
We are sincerely grateful to V. Chua, G. Fiete, Z.-C. Gu, and L. Wang for collaborations on closely related projects. H.-H. Lai thanks Kun Yang for helpful discussions in the initial stage of the work. The financial support from ARO Grant No. W911NF-09-1-0527 and NSF Grant No. DMR-0955778 for H.-H. Hung and from NSF Grant No. DMR-1004545 and No. DMR-1442366 for H.-H. Lai.


\section*{References}

\begin{thebibliography}{0}

\bibitem{fu2007}
 L. Fu and C.~L. Kane, {\it Phys. Rev. B} {\bf 76}, 045302 (2007)
 
\bibitem{moore2007}
 J.~E. Moore and L. Balents, {\it Phys. Rev. B} {\bf 75}, 121306 (2007)
 
\bibitem{moore2010}
J.~E. Moore, {\it Nature} \textbf{464}, 194 (2010)

\bibitem{roy2009}
R. Roy, {\it Phys. Rev. B} \textbf{79}, 195322 (2009)

\bibitem{hasan2010}
M.~Z. Hasan and C.~L. Kane, {\it Rev. Mod. Phys.} \textbf{82}, 3045 (2010)

\bibitem{qi2011}
X.-L. Qi and S.-C. Zhang, {\it Rev. Mod. Phys.} \textbf{83},1057 (2011)

\bibitem{Bernevig2006}
B.~A. Bernevig,  T.~L. Hughes,  and S.-C. Zhang, {\it Science} \textbf{314}, 1757 (2006)

\bibitem{Konig2007}
M. K$\ddot{o}$nig, S. Wiedmann, C. Br$\ddot{u}$ne, A. Roth, H. Buhmann, L.~W. Molenkamp, X.-L. Qi, and S.-C. Zhang, {\it Science} \textbf{318}, 766 (2007)

\bibitem{Roth:sci09}
  A. Roth, C. Br\"une, H. Buhmann, L.~W. Molenkamp, J. Maciejko, X.-L. Qi, and S.-C. Zhang, {\it Science} \textbf{325}, 294
  (2009)
  
\bibitem{Young:prb08}
M.~W. Young, S.-S. Lee, and C. Kallin, {\it Phys. Rev. B} \textbf{78},125316 (2008)
  
\bibitem{Neupert:prb11}
T. Neupert, L. Santos, S. Ryu, C. Chamon, and C. Mudry {\it Phys. Rev. B} \textbf{84}, 165107(2011)
  
\bibitem{Qi11}
 X.-L. Qi, {\it Phys. Rev. Lett.} \textbf{107}, 126803 (2011)
  
\bibitem{Levin:prl09}
 M. Levin and A. Stern, {\it Phys. Rev. Lett.} \textbf{103}, 196803 (2009)
  
\bibitem{Maciejko:prb13}
 J. Maciejko and A. R\"uegg, {\it Phys. Rev. B} \textbf{88}, 241101 (2013)
  
\bibitem{Ruegg:prl12}
 A. R\"uegg  and G.~A. Fiete, {\it Phys. Rev. Lett.} \textbf{108}, 046401 (2012)
  
\bibitem{Miguel2013}
M.~A.~N. Ara\'ujo, E.~V. Castro, and P.~D. Sacramento, {\it Phys. Rev. B} \textbf{87}, 085109 (2013)
  
\bibitem{Kargarian:prl13}
 M. Kargarian and G.~A. Fiete, {\it Phys. Rev. Lett.} \textbf{110}, 156403 (2013)
  
\bibitem{Maciejko:prl14}
 J. Maciejko, V. Chua, and G.~A. Fiete, {\it Phys. Rev. Lett.} \textbf{112}, 016404 (2014)
  
\bibitem{Pesin:np10}
D. Pesin and L. Balents, {\it Nat. Phys.} \textbf{6}, 376 (2010)
  
\bibitem{Kargarian:prb11}
 M. Kargarian, J. Wen, G.~A. Fiete, {\it Phys. Rev. B} \textbf{83}, 165112 (2011)
  
\bibitem{Maciejko:prl10}
 J. Maciejko, X.-L. Qi, A. Karch, and S.-C. Zhang, {\it Phys. Rev. Lett.} \textbf{105}, 246809 (2010)
  
\bibitem{Wan:prb11}
X. Wan, A.~M. Turner, A. Vishwanath, and S.~Y. Savrasov, {\it Phys. Rev. B} \textbf{83}, 205101 (2011)
  
\bibitem{Go:prl12}
A. Go, W. Witczak-Krempa, G.~S. Jeon, K. Park, and Y.~B. Kim, {\it Phys. Rev. Lett.} \textbf{109}, 066401 (2012)
  
\bibitem{kane2005a}
C.~L. Kane and E.~J. Mele, {\it Phys. Rev. Lett.} \textbf{95}, 226801 (2005)
  
\bibitem{haldane1988}
 F.~D.~M. Haldane, {\it Phys. Rev. Lett.} \textbf{61}, 2015 (1988)
  
\bibitem{rachel2010}
S. Rachel and K. Le Hur, {\it Phys. Rev. B} \textbf{82}, 075106 (2010)
  
\bibitem{yu2011}
 S.-L. Yu,  X.~C. Xie, and J.-X. Li, {\it Phys. Rev. Lett.} \textbf{107}, 010401 (2011)
  
\bibitem{zheng2011}
 D. Zheng, G.-M. Zhang, and C. Wu, {\it Phys. Rev. B} \textbf{84}, 205121 (2011)
  
\bibitem{hohenadler2011}
M. Hohenadler, T.~C. Lang, and F.~F. Assaad, {\it Phys. Rev. Lett.} \textbf{106}, 100403
  (2011)
  
\bibitem{Budich:prb12}
 J.~C. Budich, R. Thomale, G. Li, M. Laubach, and S.-C. Zhang, {\it Phys. Rev. B} \textbf{86}, 201407 (2012)
  
\bibitem{wuwei2012}
 W. Wu, S. Rachel, W.-M. Liu, and K. Le Hur, {\it Phys. Rev. B} \textbf{85}, 205102 (2012)
  
\bibitem{hohenadler2012}
 M. Hohenadler, Z.~Y. Meng, T.~C. Lang, S. Wessel, A. Muramatsu, and F.~F. Assaad, {\it Phys. Rev. B} \textbf{85}, 115132 (2012)
  
\bibitem{Griset2012}
 C. Griset and C. Xu, {\it Phys. Rev. B} \textbf{85}, 045123 (2012)
  
\bibitem{lang2013}
T.~C. Lang, A.~M. Essin, V. Gurarie, and S. Wessel, {\it Phys. Rev. B} \textbf{87}, 205101 (2013)
  
\bibitem{Hung2013}
 H.-H. Hung, L. Wang, Z.-C. Gu, and G.~A. Fiete, {\it Phys. Rev. B} \textbf{87}, 121113 (2013)
  
\bibitem{hung2014}
H.-H. Hung, V. Chua, L. Wang, and G.~A. Fiete, {\it Phys.
  Rev. B} \textbf{89}, 235104 (2014)
  
\bibitem{Fetter-Walecka}
A. L. Fetter and J. D. Walecka, {\it Quantum Theory of Many-Particle Systems}, Dover Publications, Inc. (2003)

\bibitem{Lai_staggerKM}
H.-H. Lai and H.-H. Hung, {\it Phys. Rev. B} \textbf{89}, 165135 (2014)

\bibitem{meng2013}
Z.~Y. Meng, H.-H. Hung, and T.~C. Lang, {\it Mod. Phys. Lett. B} \textbf{28}, 143001 (2013)

\bibitem{liujunwu}
J. Liu, T.~H. Hsieh, P. Wei, W. Duan, J. Moodera, and L. Fu, {\it Nature Materials} \textbf{13}, 178 (2014)

\bibitem{lai2014}
 H.-H. Lai, H.-H. Hung and G. A. Fiete, {\it arXiv:1407.4123} (2014)

\bibitem{white1989}
S.~R. White, D. J. Scalapino, R.~L. Sugar, E.~Y. Loh, J.~E. Gubernatis, and R.~T. Scalettar, {\it Phys. Rev. B} \textbf{40}, 506 (1989)

\bibitem{sorella1989}
 S. Sorella, S. Baroni, R. Car,  and M. Parrinello, {\it Europhys. Lett.} \textbf{8}, 663 (1989)
 
\bibitem{meng2010}
Z.~Y. Meng, T.~C. Lang, S. Wessel, F.~F. Assaad, and A. Muramatsu, {\it Nature} \textbf{464}, 847 (2010)

\bibitem{hirsch1983}
J.~E. Hirsch, {\it Phys. Rev. B} \textbf{28}, 4059 (1983)

\bibitem{assaad1998}
 F. Assaad, {\it arXiv:cond-mat/9806307} (1998)
 
\bibitem{assaad2002}
F.~F. Assaad, {\it Quantum Monte Carlo
  methods on lattices: The determinantal approach in Quantum Simulations of
  Complex Many-Body Systems: From Theory to Algorithms, Lecture Notes}, NIC Series Vol. \textbf{10} (2002)

\bibitem{WangPRB}
 Z. Wang and S.-C. Zhang, {\it Phys. Rev. B} \textbf{86}, 165116 (2012)
 
\bibitem{Wang_PRX}
Z. Wang and S.-C. Zhang, {\it Phys. Rev. X} \textbf{2}, 031008 (2012)

\bibitem{Wang2013}
 Z. Wang and B. Yan, {\it Journal of Physics: Condensed Matter} \textbf{25}, 155601 (2013)
 
 \bibitem{Shankar_RGRMP}
R. Shankar, {\it Rev. Mod. Phys.} \textbf{66}, 129--192 (1994)

\bibitem{Hohenadler_KMC}
M. Hohenadler, F. Parisen Toldin, and I. F. Herbut, and F. F. Assaad, {\it Phys. Rev. B} \textbf{90}, 085146 (2014)

\bibitem{Schuler2013}
M. Sch$\ddot{u}$ler, M. R$\ddot{o}$sner, T. O. Wehling, A. I. Lichtenstein, and M. I. Katsnelson, {\it Phys. Rev. Lett.} \textbf{111}, 036601 (2013)

\bibitem{hohenadler2013}
 M. Hohenadler, and F.~F. Assaad, {\it Journal of Physics: Condensed Matter} \textbf{25}, 143201 (2013)

\bibitem{FKM_model}
L. Fu, C. L. Kane, and E. J. Mele, {\it Phys. Rev. Lett.} \textbf{98}, 106803 (2014)

\bibitem{ojanen2013}
T. Ojanen, {\it Phys. Rev. B} \textbf{87}, 245112 (2013)


\end{thebibliography}
\end{document}